\documentclass[prd,aps,floatfix,superscriptaddress,nofootinbib,twocolumn,10pt,English]{revtex4}

\usepackage{amssymb,amsmath}
\usepackage{graphicx}
\usepackage[lofdepth,lotdepth]{subfig}
\usepackage[percent]{overpic}
\usepackage{overpic}
\usepackage{color}
\usepackage[colorlinks=true]{hyperref}
\numberwithin{equation}{section}

\newcommand{\rmin}{r_{-}}
\newcommand{\rmax}{r_{+}}

\begin{document}

\title{Relativistic Mean Motion Resonance}

\author{Huan Yang}
\affiliation{University of Guelph, Guelph, Ontario N2L3G1, Canada}
\affiliation{Perimeter Institute for Theoretical Physics, Waterloo, Ontario N2L 2Y5, Canada}
\author{B\'eatrice Bonga}
\affiliation{Perimeter Institute for Theoretical Physics, Waterloo, Ontario N2L 2Y5, Canada}
\author{Zhipeng Peng}
\affiliation{Gravitational Wave and Cosmological Laboratory, Department of Astronomy, Beijing Normal University, Beijing 100875, China}
\affiliation{Perimeter Institute for Theoretical Physics, Waterloo, Ontario N2L 2Y5, Canada}
\author{Gongjie Li}
\affiliation{Center for Relativistic Astrophysics, School of Physics, Georgia Institute of Technology, Atlanta, GA 30332, USA}

\begin{abstract}
Mean motion resonances are commonly seen in planetary systems, e.g., in the formation of orbital structure  of Jupiter's moons and the gaps in the rings of Saturn.
In this work we study their effects in fully relativistic systems. We consider a model problem with two stellar mass black holes orbiting around a supermassive
black hole. By adopting a two time-scale expansion technique and averaging over the fast varying orbital variables, we derive the effective Hamiltonian for the slowly varying
dynamical variables.
The formalism is illustrated with a $n_{\underline{\phi}} : n_{\underline{r}} : n_\phi= 2:1:-2$ resonance in Schwarzschild spacetime, which naturally becomes the $3:2$ resonance widely studied in the Newtonian limit.
We also derive the multi-body Hamiltonian in the post-Newtonian regime, where the radial and azimuthal frequencies are different because of the post-Newtonian precession.
%The impact on gravitational wave by Advection-Dominated-Accretion-Flows is of order $10^-2$, which is not detectable by LISA.
The capture and breaking conditions for these relativistic mean motion resonances are also discussed. In particular, pairs of stellar mass black holes surrounding the supermassive black hole could be locked into resonances as they enter the LISA band, and this would affect their gravitational wave waveforms. 
\end{abstract}

\maketitle

\section{Introduction}
\label{sec:intro}

Mean motion resonance is a type of orbital resonance that occurs when the orbital frequencies of two gravitationally interacting bodies, both moving in the gravitational potential well of a massive object, become commensurate with each other  \cite{Murray-Dermott}.
During these resonances, the mutual gravitational influence of the bodies is enhanced. As a result, mean motion resonances can significantly alter the orbit of one or both of the bodies.
Depending on the eccentricity and/or the relative inclination of the two bodies, they can be captured into various resonance configurations. Some of these configurations are stable, such as the orbits of Pluto and the plutinos \cite{williams1971resonances,yu1999dynamics}; and some are unstable, such as the Kirkwood gaps in the asteriod belt at $\sim 3 $ AU from our Sun \cite{wisdom1982origin,dermott1983nature}.
Stablization  may occur when the two bodies are synchronized such that they never closely approach each other, in which case the resonance is locked.
Once a pair or even a chain of objects \cite{mills2016resonant} are locked into mean motion resonance, they can migrate together towards the central massive object while keeping the ratio of orbital frequencies fixed.
The resonant locking breaks down if the adiabatic evolution of the trajectory in the phase space exits the resonance zone, or if
the external dissipative force becomes stronger than the resonant interaction between the objects so that adiabatic approximation is no longer valid.

As all previous studies of mean motion resonances are performed within the Newtonian regime, it is interesting to extend the analysis of these resonances to the relativistic setting.
A possible scenario in which relativistic effects may become important is when two stellar-mass black holes are orbiting a massive black hole. This would be a  multiple EMRI (extreme mass-ratio inspiral).
While EMRIs have not been observed yet, they are possible sources for the space-born gravitational wave detector, LISA (Laser Interferometric Space Antenna), which is scheduled to launch in 2034.
As EMRI systems generally orbit $10^4 \sim 10^5 $ cycles in the LISA band before their final plunge, any additional force or small deviation from theoretical predictions may accumulate over many cycles resulting in an amplification of the deviation. Therefore EMRIs are ideal for testing the spacetime of rotating black holes in General Relativity \cite{hughes2006sort} \footnote{However, many modified theories of gravity naturally contain coupling coefficients with negative mass dimensions due to the inclusion of higher derivative terms in the action, e.g. dynamical Chern-Simons theory \cite{alexander2009chern} and Gauss-Bonnet theory \cite{gross1987quartic}. Therefore for these theories the deviation from GR is amplified with lower-mass compact objects, and it would be preferable to use high-frequency gravitational-wave detectors \cite{miao2018towards,martynov2019exploring} instead of LISA to obtain better constraints.
}, searching for the possible existence of an ultra-light axion field \cite{zhang2019gravitational,zhang2019dynamic,hannuksela2019probing,brito2017gravitational} or other exotic matter/horizon structure \cite{cardoso2016gravitational,cardoso2017testing,maselli2018probing}, studying the physics of black hole accretion disks \cite{kocsis2011observable,barausse2014can} and the astrophysical environment of supermassive black holes in galactic centres \cite{YunesMiller2011,bonga2019tidal}.%
%%%%%beyond  EMRIs%%%%%
%Black hole spectroscopy can also be performed by observing the ringdown of massive black holes \cite{berti2006gravitational}, which may have higher signal-to-noise ratio than stellar-mass black hole binaries in the LIGO (Laser Interferometric Gravitational-Wave Observatory) band \cite{berti2016spectroscopy,yang2017black}.
%

An important difference between geodesic orbits in General Relativity (GR) and Newtonian orbits is the number of independent orbital frequencies.
Geodesics around rotating black holes naturally contain three orbital frequencies, as compared to one orbital frequency for Newtonian Keplerian orbits.
Therefore, the Newtonian condition for resonance to occur, i.e.,
\begin{equation}
	j \; \omega + \underline{j} \; \underline{\omega} \approx 0
\end{equation}
with $j, \underline{j}$ both integer and $\omega,\underline{\omega}$ describing the orbital frequency of the two bodies revolving around the massive object,
is a subspace of the resonance condition in the relativistic scenario, which is a constraint on six orbital frequencies.
In other words, the mutual gravitational interaction between different degrees of freedom allows a larger set of commensurate frequencies in GR.
As a result, we expect
%distinct resonance conditions for resonances at different order.
a richer resonant structure than in Newtonian gravity.
This is indeed the case, as we will see in Sec.~\ref{sec:GR}.

Resonant pairs are most likely to form in the Newtonian regime when both objects are far away from the massive black hole. This can happen, for instance, when both objects move within an accretion disk around the massive black hole. Such capture is similar to the standard planetary resonance capture mechanism.
After capture into resonance, these pairs will jointly migrate towards the massive black hole by gravitational wave radiation and disk dissipation.
As the resonant pair enters the strong-gravity regime, relativistic corrections start to play a role. The conservative piece of the relativistic correction only slightly changes the shape of trajectories in the phase space  diagram (Sec.~\ref{sec:post-newtonian}), where the gravitational radiation reaction tends to break the pair.
%the Newtonian resonance described by the two integers $n$ and $\underline{n}$ naturally becomes one/multiple of the relativistic mean motion resonances.
It is also possible for resonant crossing to happen, where two or more resonant conditions are approximately satisfied, so that the system jumps from one resonance to another.
If resonances start to overlap, this can lead to chaos \cite{Murray-Dermott}. We do not consider such cases here.

As the resonant pair spirals sufficiently close to the massive black hole, the gravitational radiation reaction becomes stronger than the disk force, and the resonant locking breaks down.
In this case, the inner object merges with the massive black hole first, while its orbit is still influenced by the gravitational field of the outer object. We have analyzed the effect of such scenario in a separate study \cite{bonga2019tidal}, where we show that the impact on the waveform of the inner object is possibly detectable by LISA, depending on the breaking radius of the pair. In fact, the tidal resonance effect studied there can be viewed as a ``failed" capture of the relativistic mean motion resonance proposed here.

The paper is organized as follows. In Sec.~\ref{sec:GR} we derive an effective  Hamiltonian describing the orbital dynamics near resonance in GR. We ignore any astrophysical effects and focus solely on the relativistic corrections. This effective Hamiltonian will be important in understanding the dynamics of bodies near mean motion resonance in the strong-gravity regime. Sec.~\ref{sec:resonance-example} discusses an example of mean motion resonance: two point masses moving in the equatorial plane around a non-spinning black hole, with the inner orbit circular and the outer eccentric. From the level curves of the effective Hamiltonian, we can identify the resonant and non-resonant regimes of the phase space.
In Sec.~\ref{sec:post-newtonian} we present an effective Hamiltonian describing the orbital dynamics near resonance in post-Newtonian theory  using Poincar\'e variables
and compare it to its fully relativistic counterpart.
Sec.~\ref{sec:numerics} shows the results of a numerical study in which we evolve two stellar-mass black holes orbiting within the disk around a supermassive black hole using the N-body code REBOUND and including disk effects and relativistic corrections to the conservative and disappative orbital evolution to leading post-Newtonian order. We study the capture into and breaking of resonance.
Sec.~\ref{sec:conclusion} summarizes the result and highlights some open issues.

Throughout this work we adopt geometric units by setting $c=G=1$.

\section{Relativistic Hamiltonian Formalism}\label{sec:GR}

\begin{figure}[tb]
	\includegraphics[width=6cm]{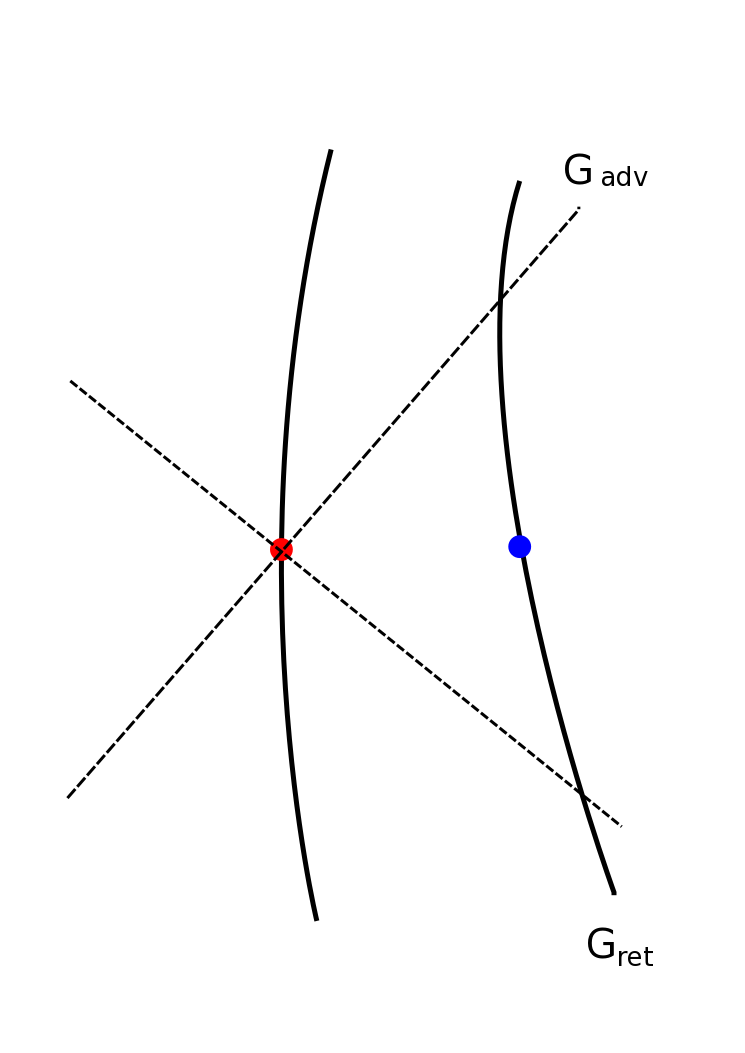}
	\caption{Illustration of the worldlines of two point masses moving under the influence of a background black hole and mutual gravitational interaction. With the conservative Hamiltonian, the motion of one point mass
		is only affect by the portion of the other point mass's worldline lying in the first mass's future and past lightcone.}
	\label{fig:worldline}
\end{figure}

We shall first consider the case that the interacting point masses are close to a massive black hole. In this fully relativistic regime,
we can adopt black hole perturbation theory, where the point masses and the gravitational fields that they generate can be treated as perturbations of the spacetime of the massive black hole.  A large body of literature has been devoted to the study of a single point mass moving in a large black hole's spacetime under the influence of its own radiation reaction (the ``self-force'' problem), which is relevant for understanding the orbital evolution of EMRIs for LISA detection\footnote{Two nice review articles on this subject are \cite{poisson2011motion,Barack:2018yvs}.}.
Here we place multiple point masses in the same background, so that their mutual gravitational interaction is also important. In addition to possible astrophysical applications, it is also theoretically interesting to explore multi-body effects in the strong-gravity regime.

\subsection{Two-body Hamiltonian in Kerr Background}
The Hamiltonian of a single point mass moving in a curved background spacetime can be written as
\begin{align}
\mathcal{H} =\frac{1}{2\mu} g_{\alpha\beta} p^\alpha p ^\beta\,.
\end{align}
With the set of canonical variables $(x^\beta, p_\beta)$ and the above Hamiltonian, one can derive the Hamilton equation of motion for this point mass\footnote{For simplicity we have taken the rest mass to be $\mu=1$.}:
\begin{align}
\frac{d x^\beta}{d \tau} = \frac{\partial \mathcal{H}}{ \partial p_\beta} =p^\beta\,, \quad \frac{d p_\beta}{d \tau} = -\frac{\partial \mathcal{H}}{\partial x^\beta}\,,
\end{align}
with $\tau$ being the proper time of the point mass.
If the background spacetime is Kerr, four conserved quantities
can be obtained from the equations of motion, and the orbital
motion becomes separable:
\begin{align}\label{eq:kerreom}
& \frac{d \theta}{d \lambda_m}
= \sqrt{Q-\cos^2\theta [a^2(\alpha^2-E^2)+L^2_z/\sin^2\theta]}\,, \nonumber \\
& \frac{d r}{d \lambda_m}
= \sqrt{[E(r^2+a^2)-L_z a]^2-\Delta [r^2+(L_z-a E)^2+Q]}\,, \nonumber \\
& \frac{d \phi}{d \lambda_m}
= -\left ( a E -\frac{L_z}{\sin^2\theta}\right )+\frac{a}{\Delta } [E(r^2+a^2)-a L_z]\,,\nonumber \\
& \frac{d t}{d \lambda_m}
= -a(a E \sin^2\theta -L_z) +\frac{r^2+a^2}{\Delta} [E(r^2+a^2)-L_z a]\,.
\end{align}
where $M,a$ are the mass and spin of the Kerr black hole, $E, Q, L_z$ are the point mass energy, Carter's constant and angular momentum along the symmetry axis, respectively. $\Delta$ is defined as $r^2-2M r +a^2$ and Mino time $\lambda_m$ is related to proper time by $\lambda_m$  $d/d\lambda_m =(r^2+a^2\cos^2 \theta) d/d\tau$.

The stellar-mass objects of interest in this study are (of course) not point particles. If they carry nonzero quadrupole moments and other higher order moments, additional couplings with the background curvature are expected. These additional complexities may be neglected if the stellar-mass object is a Schwarzschild black hole and its size is much smaller than the background radius of curvature. However, even in the point mass limit, in principle one should take into account the interaction of the object with its own gravitational field. This is known as the gravitational self-force.
In order to derive the conserved dynamics of a point mass under the influence of the conservative piece of the self-force, one can decompose the metric as $g_{\alpha\beta} =g^{\rm Kerr}_{\alpha\beta} +h_{\alpha\beta}$ \cite{poisson2011motion}, with the  metric perturbation given by
\begin{align}\label{eq:hgreen}
h_{\alpha\beta}(x) =\mu \int d \tau' \,G_{\alpha\beta\rho\sigma} (x ; x') {u'}^\rho {u'}^\sigma\,.
\end{align}
Here $G$ is the half retarded, half advance Green function:
\begin{align}
G =\frac{1}{2}(G_{\rm ret} +G_{\rm adv})\,,
\end{align}
which is symmetric under time reversal operation ($t \rightarrow -t$).
This metric perturbation $h_{\alpha\beta}$ has a diverging part, and one
has to subtract a singular piece.  The detailed procedure, known as {\it regularization}, is nicely explained in \cite{barack2009gravitational}.

For the purpose of this study, however, the conservative self-interactions of the stellar-mass objects are unimportant and shall be neglected hereafter, as they do not
contribute to the resonance terms \footnote{They may affect the sustained transient resonance described in \cite{van2014conditions} for a single EMRI object.}. However, the mutual gravitational interaction is important. We write the total Hamiltonian of two interacting bodies as
\begin{align}\label{eq:htot}
\mathcal{H} = & \, \frac{\mu}{2} g^{\rm Kerr}_{\alpha\beta}(x) u^\alpha u^\beta + \frac{\underline{\mu}}{2} g^{\rm Kerr}_{\alpha\beta}(\underline{x}) \underline{u}^\alpha\underline{ u}^\beta \nonumber \\
& + \frac{\mu}{2} \underline{h}_{\rm \alpha\beta}(x) u^\alpha u^\beta + \frac{\underline{\mu}}{2} h_{\rm \alpha\beta}(\underline{x}) \underline{u}^\alpha\underline{ u}^\beta\, \nonumber \\
: = & \,\mathcal{H}_0 + \epsilon \; \mathcal{H}_{\rm int}\,
\end{align}
where $\mathcal{H}_0$ is the unperturbed Hamiltonian in the first line, $\mathcal{H}_{\rm int}$ is the perturbation described in the second line and $\epsilon$ is a book-keeping index for the perturbative Hamiltonian.
The $x, u$ are the position and velocity of the inner point mass, and $\underline{x}, \underline{u}$ are those of the outer point mass.
The metric perturbation $h, \underline{h}$ can be obtained from Eq.~\eqref{eq:hgreen} by plugging in the worldliness of the inner and outer point masses, respectively.

It is important to note that, at any given time, the inner object is only influenced by the part of outer object's worldline within the future and past lightcone of the inner object, and vice versa (c.f. Fig.~\ref{fig:worldline}).
This means that $\underline{h}_{\alpha \beta}(x)/h_{\alpha \beta}(\underline{x})$ is in principle independent of the outer/inner object's motion at time $t$, as they are causally disconnected.
However, it turns out that it is still possible to write $\underline{h}_{\alpha \beta}(x)/h_{\alpha \beta}(\underline{x})$ as functions of $\underline{x} /x$ at any given time.
This is because to the leading order in mass ratios ($\eta:=\mu/M, \eta':=\mu'/M$), Eq.~\eqref{eq:hgreen} only depends on the unperturbed, geodesic orbit of the point mass that sources the gravitational field \footnote{Similar observations have been made for computing the leading order gravitational self-force of EMRIs.}.
As the geodesic orbit is deterministic, specifying the position and momentum at any instant in time determines the whole worldline, including the parts that extend to the other point mass's past and future lightcones.

\subsection{Canonical transformations}

To analyze this problem in detail, we first separate the average adiabatic, the resonant and the non-resonant oscillatory terms in the Hamiltonian $\mathcal{H}$ using action-angle variables $\{q, J\}, \{\underline{q}, \underline{J}\}$.
The derivation that follows is rather general. In particular, it does not rely on the explicit form of the interaction Hamiltonian in Eq.~\ref{eq:htot}, to which we return  in the next section.

As the generalized angles $q, \underline{q}$ have a period of $2\pi$, the perturbed Hamiltonian can be simply decomposed as a Fourier series ($j=r,\theta,\phi$):
\begin{align}
\mathcal{H}_{\rm int} = \sum_{n_j, \underline{n}_j} H_{n_j, \underline{n}_j} (J, \underline{J})e^{i  n_j q_j + i\underline{n}_j \underline{q}_j } \,,
\end{align}
with the Fourier coefficients given by
\begin{align}\label{eq:hnnp}
H_{n_j, \underline{n}_j} &= \frac{1}{(2\pi)^6} \int d^3 q_j d^3 \underline{q}_j \frac{\mu}{2}\underline{h}_{\alpha \beta} (x; \underline{x}) u^\alpha u^\beta  e^{-i  n_j q_j - i\underline{n}_j \underline{q}_j } \nonumber \\
& +  \frac{1}{(2\pi)^6} \int d^3 q_j d^3 \underline{q}_j  \frac{\underline{\mu}}{2} h_{\rm \alpha\beta}(\underline{x};x) \underline{u}^\alpha\underline{ u}^\beta e^{-i  n_j q_j - i\underline{n}_j \underline{q}_j } \,.
\end{align}
Mean motion resonance may happen if a certain combination of $q, \underline{q}$, specifically $ n_j q_j +\underline{n}_j \underline{q}_j $ with $n_j$ and $\underline{n}_j$ being integer, becomes a slowly varying quantity \footnote{The Einstein summation rule has been applied, so that the explicit summation symbol is abbreviated.}. In other words, when  $n_j \omega_j +\underline{n}_j \underline{\omega}_j \approx 0$ is satisfied, with $\omega_j, \underline{\omega}_j$ being the frequencies of motion in $r, \theta,\phi$ directions.
Using this observation, the interaction Hamiltonian can be separated into an average adiabatic term, resonant terms and non-resonant oscillating terms:
\begin{align}
\mathcal{H}_{\rm int} =& \bar{H}(J, \underline{J}) + \sum_{k} H_{k N_j, k\underline{N}_j} (J, \underline{J})e^{i k N_j q_j + i k\underline{N}_j \underline{q}_j } \nonumber \\
&+ \sum_{n_j, \underline{n}_j \in R} H_{n_j, \underline{n}_j} (J, \underline{J})e^{i  n_j q_j + i\underline{n}_j \underline{q}_j }\,,
\end{align}
where on resonance $n_j = N_j, \underline{n}_j = \underline{N}_j$, $k$ is a non-zero integer and $R$ is the set of all non-resonant 6-tuples: $\{(n_j, \underline{n}_j) \in \mathbb{Z} | (n_j, \underline{n}_j) \neq k (N_j, \underline{N}_j) \; \forall \; k \in \mathbb{Z}^{\neq} \} $.
While the Hamiltonian nicely separates, the equations of motion for the action-angle variables couple oscillatory and resonant terms:
\begin{align}
\frac{d q_i}{d \tau}
&= \frac{\partial \mathcal{H}}{\partial J^i} \nonumber \\
&= \Omega_i + \epsilon \frac{\partial \bar{H}(J,\underline{J})}{\partial J^i} \nonumber \\
&\quad
+ \epsilon \sum_k \frac{\partial H_{k {\bf  N}, k {\bf  \underline{N}}}}{\partial J^i}e^{i  k N_j q_j + i k \underline{N}_j \underline{q}_j }  \nonumber \\
& \quad
+ \epsilon \sum_{n_j, \underline{n}_j \in R} \frac{\partial H_{n_j, \underline{n}_j} }{\partial J^i} e^{i  n_j q_j + i\underline{n}_j \underline{q}_j }   \\
\frac{d J_i}{d \tau}
&= - \frac{\partial \mathcal{H}}{\partial q^i} \nonumber \\
&=\,-i\,\epsilon   \sum_k k \; N_i \; H_{k {\bf  N}, k {\bf  \underline{N}}} e^{i  k N_j q_j + i k\underline{N}_j \underline{q}_j }
\,,
\end{align}
where $\Omega_i :=\partial \mathcal{H}_0/\partial J^i$ and analogous equations hold for $\underline{q}_i, \underline{J}_i$. We introduce a change of variables to eliminate the dependence on the rapidly oscillating non-resonant terms to order $\epsilon$ following techniques in \cite{kevorkiancole}. This change of variables is known as a `near-identity' transformation because for $\epsilon=0$ this reduces to the identity transformation:
\begin{align}
\tilde{q}_i (\vec{q}, \vec{J}, \vec{\underline{q}}, \vec{\underline{J}}) = q_i + \epsilon \; L_i(\vec{q}, \vec{J}, \vec{\underline{q}}, \vec{\underline{J}}) + \mathcal{O}(\epsilon^2) \\
\tilde{J}_i (\vec{q}, \vec{J}, \vec{\underline{q}}, \vec{\underline{J}}) = J_i + \epsilon \; T_i(\vec{q}, \vec{J}, \vec{\underline{q}}, \vec{\underline{J}}) + \mathcal{O}(\epsilon^2)
\end{align}
where
\begin{align}
L_i &= \bar{L}_i(\vec{J},\vec{\underline{J}})  \nonumber  \\
&\qquad + i \sum_{n_j, \underline{n}_j \in R} \left[ \frac{1}{n_j \Omega_j+\underline{n}_j \underline{\Omega}_j} \right]\frac{\partial H_{n_j, \underline{n}_j} }{\partial J^i}  e^{i  n_j q_j + i\underline{n}_j  \underline{q}_j }  \label{eq:Li}  \\
T_i & = \bar{T}_i(\vec{J},\vec{\underline{J}})    \nonumber \\
&\qquad+ i \sum_{n_j, \underline{n}_j \in R}  \left[\frac{n_i}{n_j \Omega_j+\underline{n}_j \underline{\Omega}_j}   \right] H_{n_j, \underline{n}_j} e^{i  n_j q_j + i\underline{n}_j  \underline{q}_j } \; , \label{eq:Ti}
\end{align}
with $\bar{L}_i$ and $\bar{T}_i$ arbitrary functions of $J_i$ and $\underline{J}_i$. The underlined variables are analogously transformed. The freedom in $\bar{L}_i$ and $\bar{T}_i$  can be used to further simplify the equations, which may be convenient as one considers higher orders in $\epsilon$; for the calculation at hand, their behavior is irrelevant.
The new variables now satisfy:
\begin{align}
& \frac{d \tilde{q}_i}{d \tau}
=\Omega_i + \epsilon \frac{\partial \bar{H}}{\partial \tilde{J}^i} + \epsilon \sum_k  \frac{\partial H_{k {\bf  N}, k {\bf  \underline{N}}}}{\partial \tilde{J}^i}e^{i  k N_j \tilde{q}_j + i k \underline{N}_j \tilde{\underline{q}}_j } + \mathcal{O}(\epsilon^2)\,, \\
&\frac{d \tilde{J}_i}{d \tau}
=\, -i \,\epsilon \sum_k k \;  N_i\;  H_{k {\bf  N}, k {\bf  \underline{N}}}e^{i  k N_j \tilde{q}_j + i k \underline{N}_j \tilde{\underline{q}}_j }+ \mathcal{O}(\epsilon^2) \,, \nonumber \\
\end{align}
where the non-resonant terms now decouple  and do not contribute to the secular evolution of $\{ \tilde{q}_i, \tilde{J}_i\}$ near resonance. Similar transformations and resulting equations hold for $\{ \tilde{\underline{q}}_i, \tilde{\underline{J}}_i\}$.

%To make it obvious that the terms on the right hand sides of the equations of motions for $\{ \tilde{q}_i, \tilde{J}_i\}$ only depend on the resonant $\tilde{q}_i$ and $\tilde{\underline{q}}_i$,
In order to summarize the above equations of motion into one that is only controlled by the resonant degree of freedom,
we define
\begin{align}\label{eq:Qsum}
Q(\lambda) := \sum_i N_i \tilde{q}_i \left (\int \Gamma_i d\lambda \right ) + \sum_i \underline{N}_i \tilde{\underline{q}}_i \left (\int \underline{\Gamma}_i d\lambda \right )\,,
\end{align}
with $\Gamma_i := \omega_i/\Omega_i, \underline{\Gamma}_i := \underline{\omega}_i/\underline{\Omega}_i$. Similar to $\Omega, \underline{\Omega}$, the functional dependence of $\omega, \underline{\omega}$ on $J, \underline{J}$ can be obtained from the geodesic motion. While the action-angle variables depend only on their local proper time, physically $\lambda$ plays the role of the coordinate time, such that the angles in Eq.~\eqref{eq:Qsum}  can add up with the same $\lambda$. With the presence of mutual gravitational interaction and near resonance, we can further define a {\it slow time} $\hat{\lambda} :=\epsilon \lambda$.
%and further expand various quantities as
%\begin{align}
%& J_i = J^0_i +\epsilon J^1_i +\mathcal{O}(\epsilon^2) \,,\nonumber \\
%& \omega_i = \omega^0_i +\sqrt{\epsilon} \omega^1_i +\mathcal{O}(\epsilon)\,, \nonumber \\
%& \Omega_i = \Omega^0_i +\epsilon \Omega^1_i +\mathcal{O}(\epsilon^2)\,,
%\end{align}
%and similarly for $\underline{J}, \underline{\Omega}$.
So that the equations of motion become
\begin{align}\label{eq:eom}
\frac{d Q}{d \hat{\lambda}} & = \frac{1}{\epsilon}\sum_i ( N_i \omega_i + \underline{N}_i \underline{\omega}_i )
+ \sum_i \Gamma_i N_i \frac{\partial \bar{H}}{\partial \tilde{J}^i} \nonumber \\
& \quad
+ \sum_i \underline{\Gamma}_i \underline{N}_i \frac{\partial \bar{H}}{\partial \tilde{\underline{J}^i}}
+\sum_{i,k} \Gamma_i N_i\frac{\partial H_{k {\bf  N}, k {\bf  \underline{N}}}}{\partial J_i} e^{i k Q}\nonumber \\
& \quad
+\sum_i\underline{\Gamma}_i \underline{N}_i \sum_k \frac{\partial H_{k {\bf  N}, k {\bf  \underline{N}}}}{\partial \underline{J}_i} e^{i k Q}+\mathcal{O}(\epsilon)\,,
\nonumber \\
% & \approx \sum_i \Gamma^0_i N_i \left [ \frac{\partial \Omega_i}{\partial J_j} J^1_j  +\sum_k \frac{\partial H_{k {\bf  N}, k {\bf  \underline{N}}}}{\partial J_i} e^{i k Q}\right ] \nonumber \\
%&+\sum_i\underline{\Gamma}^0_i \underline{N}_i \left [\frac{\partial \underline{\Omega}_i }{\partial \underline{J}_j}\underline{J}^1_j  +\sum_k \frac{\partial H_{k {\bf  N}, k {\bf  \underline{N}}}}{\partial \underline{J}_i} e^{i k Q}\right ]
\nonumber \\
%& \approx  \sum_{ij} \Gamma^0_i N_i  \frac{\partial \Omega_i}{\partial J_j} J^1_j +\underline{\Gamma}^0_i  \underline{N}_i \frac{\partial \underline{\Omega}_i }{\partial \underline{J}_j}\underline{J}^1_j\,,\nonumber \\
\frac{d \tilde{J}_i}{d\hat{\lambda}} & =-i\, \sum_k k \Gamma_i N_i H_{k {\bf  N}, k {\bf  \underline{N}}} e^{i  k Q} +\mathcal{O}(\epsilon) \,,\nonumber \\
\frac{d \tilde{\underline{J}_i}}{d\hat{\lambda}} & =-i\, \sum_k k \underline{\Gamma}_i  \underline{N}_i H_{k {\bf  N}, k {\bf  \underline{N}}} e^{i k Q} +\mathcal{O}(\epsilon) \,.
\end{align}
Here, we focus on the system of equations for $Q, \tilde{J}_i, \tilde{\underline{J}_i}$; the other phases not encoded in $Q$ can be recovered by direct integration after having solved for $Q, \tilde{J}_i, \tilde{\underline{J}_i}$.
We denote $\Delta \omega: =  \frac{1}{\epsilon}\sum_i ( N_i \omega_i +\underline{N}_i \underline{\omega}_i )$ and note that resonance is only present if this quantity is proportional to the mass ratio between the point masses and the primary massive black hole. This observation is similar to the analysis of mean motion resonance in the Newtonian limit \cite{Murray-Dermott}. In other words, in order to find resonance, this term is comparable to the terms involving $H_{k {\bf  N}, k {\bf  \underline{N}}} $, which is also proportional to the mass ratio. In principle with the initial condition defined, Eq.~\eqref{eq:eom} is able to predict the system evolution at any later time (provided of course all the above approximations still hold).

To understand the long-term dynamics of the mean motion resonance,  it is often convenient to use a simplified, effective Hamiltonian and study its level sets.
However, based on Eq.~\eqref{eq:eom}, it is not clear whether it is possible to write down an effective Hamiltonian for $Q$ and its conjugate momentum $\Theta$ in the most general setting. We can nevertheless restrict to the parameter regime where $\Gamma_i, \Gamma'_i$ are approximately constants.
As they are functions of action variables, they are approximately constants whenever resonant motion only induces small variation on the conserved quantities, e.g., the cases with eccentricity being small.
We will assume this is the case.
We also note that the terms coming from the adiabatic part of the Hamiltonian, while generically important to understand the dynamics of the system, can be regarded as constants near resonance and we shall drop them from hereon after.

Under these approximations, let us expand $\Delta \omega$:
\begin{align}
\Delta \omega = \left . \Delta \omega \right  |_{\Theta=0} + \left . \frac{\partial \Delta \omega}{\partial \Theta} \right  |_{\Theta=0} + \mathcal{O}(\epsilon)
\end{align}
and rewrite the first line in Eq.~\eqref{eq:eom} as
\begin{align}\label{eq:eomepan}
\frac{d Q}{d \hat{\lambda}} & \approx \alpha + 2\beta \Theta+\sum_{i,k} \Gamma_i N_i\frac{\partial H_{k {\bf  N}, k {\bf  \underline{N}}}}{\partial J_i} e^{i k Q}\nonumber \\
&+\sum_i\underline{\Gamma}_i \underline{N}_i \sum_k \frac{\partial H_{k {\bf  N}, k {\bf  \underline{N}}}}{\partial \underline{J}_i} e^{i k Q}+h.c.\,,
\end{align}
with
\begin{align}
& \alpha :=\left . \Delta \omega \right  |_{\Theta=0} \,,\nonumber \\
& \beta : = \frac{1}{2}\left . \frac{\partial \Delta \omega}{\partial \Theta} \right  |_{\Theta=0}\,.
\end{align}
Given that $\Delta \omega = \Delta \omega (\Theta, \hat{\lambda})$, the associated Hamiltonian is non-linear in $\Theta$ and this allows for the mean motion resonance to occur. If $\Delta \omega$ was independent of $\Theta$, only transient resonance would be possible.
We make the following identification with the action variables:
\begin{align}
& N_i \Gamma_i \Theta+\Theta_i = \tilde{J}_i,  \nonumber \\
& {\underline{N}_i \underline{\Gamma}_i \Theta +\underline{\Theta}_i = \underline{\tilde{J}}_i}\,,
\end{align}
with $\Theta_i, \underline{\Theta}_i$ being constants, so that Eq.~\eqref{eq:eomepan} and Eq. ~\eqref{eq:eom} are compatible with the effective Hamiltonian
\begin{align}\label{eq:ham}
{ H}_{\rm eff} = & \alpha \Theta +\beta \Theta^2 +\sum_k H_{k {\bf  N}, k {\bf  \underline{N}}} e^{i k Q } \nonumber \\
= & \, \alpha \Theta +\beta \Theta^2 +2 \sum_{k\geq 1} {\rm Re} (H_{k {\bf  N}, k {\bf  \underline{N}}} ) \cos { k Q } \nonumber \\
& \, -2\sum_{k\geq 1} {\rm Im} (H_{k {\bf  N}, k {\bf  \underline{N}}} ) \sin { k Q }\,.
\end{align}
%These equations of motion also imply that
%\begin{align}\label{eq:2ndeom}
%&\frac{d^2 Q}{d \hat{\lambda}^2} - 2\sum_{ij}\left (\Gamma^0_i \frac{\partial \Omega_i}{\partial J_j} N_i N_j+\underline{\Gamma}^0_i \frac{\partial \underline{\Omega}_i}{\partial \underline{J}_j} \underline{N}_i \underline{N}_j \right ) \nonumber \\
%&\times \sum_k k [{\rm Im}(H_{k {\bf  N}, k {\bf  \underline{N}}}) \cos {kQ}+{\rm Re}(H_{k{\bf N}, k {\bf  \underline{N}}}) \sin {k Q}]=0\,.
%\end{align}
%Similar to the discussion in [] for sustained transient resonance, we shall define an effective Hamiltonian as
%\begin{align}\label{eq:heff}
%H_{\rm eff} : & = \frac{1}{2}  ( P)^2+ 2\sum_{ij}\left (\Gamma^0_i \frac{\partial \Omega_i}{\partial J_j} N_i N_j+\underline{\Gamma}^0_i \frac{\partial \underline{\Omega}_i}{\partial \underline{J}_j} \underline{N}_i \underline{N}_j \right ) \nonumber \\
%&\times \sum_k  [-{\rm Im}(H_{k {\bf  N}, k {\bf  \underline{N}}}) \sin {kQ}+{\rm Re}(H_{k{\bf N}, k {\bf  \underline{N}}}) \cos {k Q}]
%\end{align}
%with $P = d Q/d \hat{\lambda}$. Such effective Hamiltonian reproduces Eq.~\eqref{eq:2ndeom}, and can be used for analyze the phase space of resonant motion, as discussed in Sec. III and Sec. IV.

In order to study the resonance dynamics described by this effective Hamiltonian, it is necessary to explicitly write down the dependence of $H_{ {\bf  N},  {\bf  \underline{N}}}$ on $\Theta$. As we shall see in Sec.~\ref{sec:post-newtonian}, in the Newtonian limit, the interaction Hamiltonian scales as $\Theta^{1/2}$ for the lowest order resonances (that is, those with $n_r, n_{\underline{r}} = \pm 1$), and $\Theta^{N/2}$ with $N \ge 1$ in general. In the relativistic regime, we shall assume that similar power-law behavior still holds when $\Theta$ is small. For example, for the resonance considered in Sec.~\ref{sec:resonance-example}, it is natural to expect that $\Theta \propto \underline{J}_r \propto \underline{e}^2$, with $e$ being the eccentricity, and $H_{ {\bf  N},  {\bf  \underline{N}}} \propto \underline{e}$ when $\underline{e} \ll 1$. As a result, we expect $H_{ {\bf  N},  {\bf  \underline{N}}} \propto \Theta^{1/2}$.
\\

{\it Remark.}
The above canonical transformations from the action-angle variables to the final $\{Q,\Theta \}$ can be summarized as follows:
\begin{equation}
\{q_i,J_i,\underline{q}_i,\underline{J}_i\} \stackrel{F_1}{\longrightarrow} \{\tilde{q}_i,\tilde{J}_i,\tilde{\underline{q}}_i,\tilde{\underline{J}_i}\} \stackrel{F_2}{\longrightarrow} \{Q,\Theta \}
\end{equation}
where $F_1$ and $F_2$ are generating functions, given by:
\begin{align}
F_1(q, \tilde{J}, \underline{q}, \tilde{\underline{J}})&= \sum_{i=1}^3 \left( q_i \; \tilde{J}_i + \underline{q}_i \underline{\tilde{J}_i}  \right) + \epsilon F_1^{(1)}(q, \tilde{J}, \underline{q}, \tilde{\underline{J}}) \notag \\
F_2(\tilde{q},\Theta) &= \sum_k \left(N_k \tilde{q}_k + \underline{N}_k \tilde{\underline{q}}_k \right) \Theta + \ldots
\end{align}
with the dots indicating (irrelevant) non-resonant terms and $L_i$ and $T_i$ in Eqs.\eqref{eq:Li}-\eqref{eq:Ti} are related to $F^{(1)}$ through its derivatives
\begin{align}
 L_i &= \frac{\partial F^{(1)}}{\partial \tilde{J}^i} \\
 T_i &= - \frac{\partial F^{(1)}}{\partial q^i} \; .
\end{align}
Canonical transformations of course do not reduce the size of phase space, the fact that we go from $2 \times 6$ to only $1 \times 2$ variables is due to the fact that we have decoupled the oscillatory pieces from the resonant ones and we focus only on the behavior of the resonant terms.

\section{Resonance example}
\label{sec:resonance-example}
In this section, we illustrate how to compute the resonance Hamiltonian, by explicitly evaluating an example  of a $n_{\underline{\phi}} : n_{\underline{r}} : n_\phi= 2:1:-2$ resonance with system parameters given in Table.~\ref{tab:orbit}.

\subsection{General prescription for calculating the interaction Hamiltonian}
In Sec.~\ref{sec:GR} we have derived the effective Hamiltonian of two point masses undergoing relativistic mean motion resonance, as shown in Eq.~\eqref{eq:ham}.
The key part of this effective Hamiltonian is the interaction part $H_{{\bf N, N'}}$, which requires Fourier transforming  the metric perturbation generated by a moving point mass.
Let us consider the part of the integral in Eq.~\eqref{eq:hnnp},
\begin{align}
h^N_{\alpha \beta} (\underline{x} ) :=\frac{1}{(2\pi)^3} \int d^3 {\bf q}\, h_{\alpha \beta}(\underline{x} ; x) e^{-i {\bf N} \cdot {\bf q}}\,,
\end{align}
with the inverse decomposition given by
\begin{align}
h_{\alpha \beta}(\underline{x} ; x) = \sum_{\bf N} h^N_{\alpha \beta} (\underline{x} ) e^{i {\bf N} \cdot {\bf q}}\,.
\end{align}
For a fixed worldline of the source, the metric perturbation at different $\underline{t}$ undergoes periodic oscillations, as the source is periodic. In other words, we can also write
\begin{align}
h_{\alpha \beta}(\underline{x} ; x) =\sum_{\bf N} h_{{\bf \omega N};\alpha,\beta} (\underline{r},\underline{\theta},\underline{\phi}) e^{-i {\bf \omega} \cdot {\bf N} \underline{t}}
\end{align}
with ${\bf \omega} = (\omega_r, \omega_\theta,\omega_\phi)$, and the inverse transformation
\begin{align}\label{eq:ho1}
h_{{\bf \omega N};\alpha,\beta} (\underline{r},\underline{\theta},\underline{\phi}) = {\lim_{T \rightarrow \infty}}\frac{1}{T}\int^T_0 d\underline{t} \,h_{\alpha \beta}(\underline{x} ; x) e^{i {\bf \omega} \cdot {\bf N} \underline{t}}\,.
\end{align}

Now the time translational invariance implies that (with $x_0=\{t_0,r_0,\theta_0,\phi_0\}$, $\underline{x} =\{\underline{t}, \underline{r},\underline{\theta},\underline{\phi} \}$)
\begin{align}
 h_{\alpha \beta}(\underline{t}+t, \underline{r},\underline{\theta},\underline{\phi} ;x_0)
 = h_{\alpha \beta}(\underline{x}; t_0-t, r_0,\theta_0, \phi_0) \,.
\end{align}
As a result,  Eq.~\eqref{eq:ho1} can be rewritten as
\begin{align}\label{eq:ho2}
& h_{{\bf \omega \cdot N};\alpha,\beta} (\underline{r},\underline{\theta},\underline{\phi})  = {\lim_{T \rightarrow \infty}}\frac{1}{T}\int^T_0 dt\, h_{\alpha \beta}(\underline{x} ; x)|_{\underline{t}=0} e^{-i {\bf \omega} \cdot {\bf N} t}\,\nonumber \\
 & = {\lim_{T \rightarrow \infty}}\frac{1}{T}\int^T_0 dt\, \sum_{\bf N'} h^{N'}_{\alpha \beta} (\underline{x} ) |_{\underline{t}=0}e^{i {\bf N'} \cdot {\bf q}} e^{-i {\bf \omega} \cdot {\bf N} t}\,.
\end{align}
The above expression can be re-casted in a simpler form:
\begin{align}
h_{{\bf \omega \cdot N};\alpha,\beta} =M_{\bf N, N'}\,h^{N'}_{\alpha \beta}
\end{align}
or
\begin{align}\label{eq:exp}
h^{N'}_{\alpha \beta}  =M^{-1}_{\bf N, N'}\, h_{{\bf \omega \cdot N};\alpha,\beta}
\end{align}
with (${\bf \Omega} = (\Omega_r, \Omega_\theta,\Omega_\phi)$)
\begin{align}
M_{\bf N, N'}\ : ={\lim_{T \rightarrow \infty}} \frac{1}{T}\int^T_0 d\tau u^t \,  e^{i {\bf N'} \cdot {\bf \Omega} \tau} e^{-i {\bf \omega} \cdot {\bf N} t(\tau)}\,.
\end{align}

In reality, $h_{{\bf \omega \cdot N};\alpha,\beta} $ can be obtained from a frequency-domain code that computes the metric perturbation, or reconstructed from master variables (such as Teukolsky variables or master variables in the Regge-Wheeler equation) in a frequency-domain code. Eq.~\eqref{eq:exp} then enables us to compute $H_{{\bf N, N'}}$ from $h_{{\bf \omega \cdot N};\alpha,\beta} $.

\subsection{Frequency-domain Schwarzschild metric perturbation}

\begin{table}
	\begin{tabular}{|c|c|c|c|c|c|}
		\toprule
		$\underline{r}_{\rm min}/M$ & $\underline{r}_{\rm max}/M$ & $r/M$ & $\underline{\omega}_\phi M$  & $\underline{\omega}_r M$  & $\omega_\phi M$ \\
		\hline
		%${\rm I}$ & 8 & 9 & 7.237 & 0.04044 & 0.02183 & 0.05136\\
		%\hline
		25 & 30 & 21.53 & $6.94\times 10^{-3}$ & $6.13\times 10^{-3}$ & $10^{-2}$ \\
		\hline
	\end{tabular}
		\caption{ Orbitals parameters for the resonant, equatorial orbit studied in Sec.~\ref{sec:resonance-example}. This orbit corresponds to the case with $2 \, \underline{\omega}_\phi + \underline{\omega}_r -2 \,\omega_\phi \approx 0$.
		\label{tab:orbit}}
\end{table}

\begin{figure*}
	\subfloat{
		\includegraphics[width=8cm]{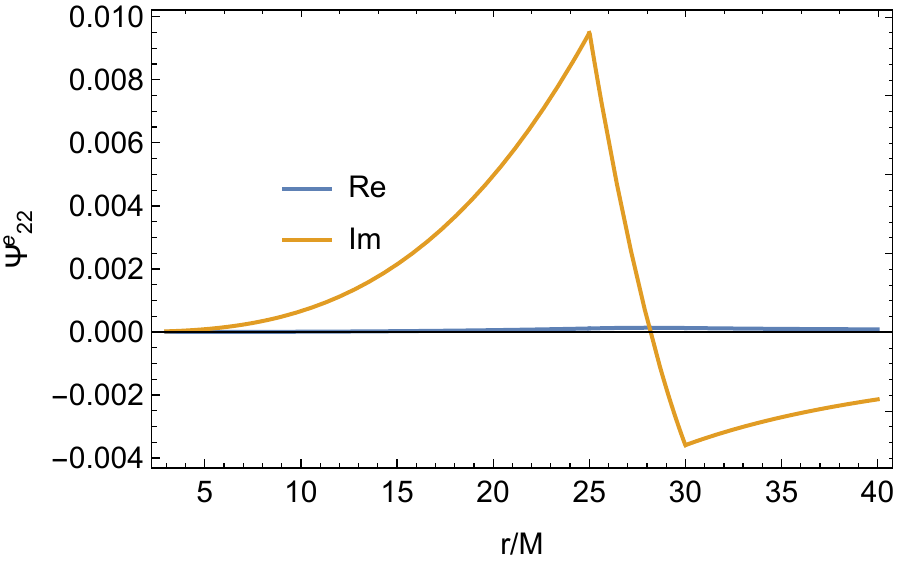}
	}
	\subfloat{
		\includegraphics[width=8cm]{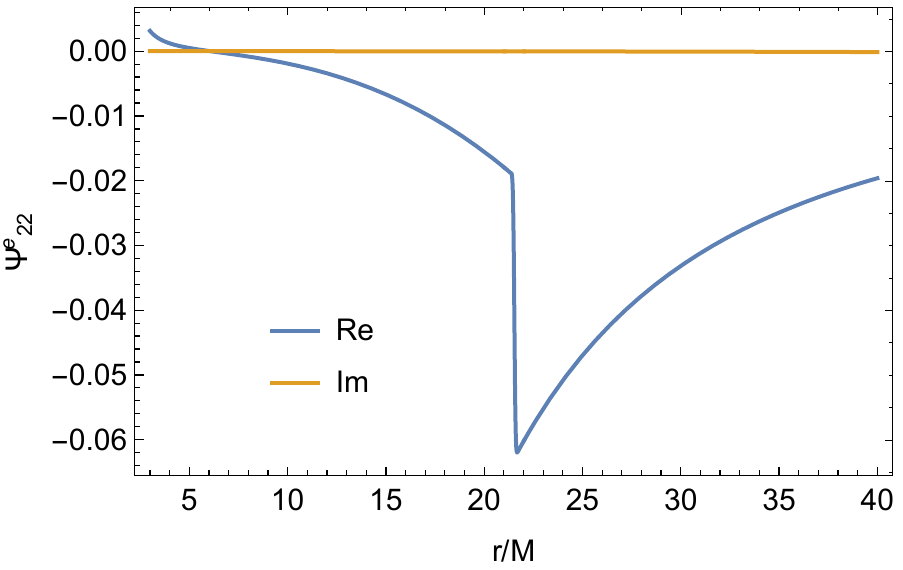}
	}
	\caption{The even parity master variable as a solution of the $\ell=2, m=2$ Master equation Eq.~\eqref{eq:mee} with eccentric (Top panel) and circular (Bottom panel) source term as described in Table \ref{tab:orbit}. The odd parity source terms are zero in this case, so the odd parity metric perturbations are also zero.}
	\label{fig:psi}
\end{figure*}

For simplicity, let us illustrate relativistic mean motion resonance in the Schwarzschild spacetime, with two point masses moving along equatorial orbits.
At the leading order, i.e. in the geodesic limit, we assume that the inner point mass moves along a circular orbit and the outer point mass moves along an eccentric orbit.
We assume that the system is close to the resonance such that
\begin{align}
2 \underline{\omega}_{\phi} +\underline{\omega}_r -2\omega_\phi \approx 0\,.
\end{align}
In other words, we shall consider the dynamical variable $Q =2 \underline{q}_\phi + \underline{q}_r - 2 q_\phi$\,.
This can be  achieved with a range of possibilities, and we will  adopt the values shown in Table \ref{tab:orbit} for constructing the point mass trajectory.
Notice that in the Newtonian limit $\omega_\phi$ becomes similar to $\omega_r$, so that the $n_{\underline{\phi}} : n_{\underline{r}} : n_\phi= 2:1:-2$ resonance considered here naturally becomes the $3:2$ outer resonance
well studied in planetary systems.
% In fact, it is possible to have other types of resonances, such as $2\underline{q}_\theta +\underline{q}_r -2 q_\phi$, that asymptote the same $3:2$ resonance
%in the Newtonian limit. It can be viewed as a degenerate level in the Newtonian regime bifurcates into distinct levels in the relativistic limit.

In the frequency domain, the metric perturbation of Schwarzschild black holes, decomposed as spherical harmonics, can be written  as (following the convention in \cite{sago2003gauge})
\begin{widetext}

\begin{align}
{\bf h}_{\ell m} =& f(r) H_{0 \ell m}(r) {\bf a}^{(0)}_{\ell m}+H_{1\ell m}(r) {\bf a}^{(1)}_{\ell m} +\frac{1}{f(r)} H_{2\ell m}(r) {\bf a}_{\ell m} + h^{(e)}_{0 \ell m}(r) {\bf b}^{(0)}_{\ell m}+ h^{(e)}_{1 \ell m}(r) {\bf b}_{\ell m} \nonumber \\
& + \frac{r^2}{2} G_{\ell m}(r) {\bf f}_{\ell m} +r^2 \left [ K_{\ell m} (r) -\frac{\ell(\ell+1)}{2} G_{\ell m}(r)\right ] {\bf g}_{\ell m}
-h_{0 \ell m}(r) {\bf c}^{(0)}_{\ell m} - h_{1 \ell m}(r) {\bf c}_{\ell m}
+ i h_{2 \ell m}(r) {\bf d}_{\ell m}\,,
\end{align}
\end{widetext}
where $f(r) = 1 -2M/r$, the tensor components ${\bf a}^{(0)}_{\ell m}, {\bf a}^{(1)}_{\ell m}, {\bf a}_{\ell m}, {\bf b}^{(0)}_{\ell m}, {\bf b}_{\ell m}, {\bf f}_{\ell m}, {\bf g}_{\ell m}, {\bf c}^{(0)}_{\ell m},  {\bf c}_{\ell m}, {\bf d}_{\ell m}$ are given in Appendix \ref{sec:tensor} and the common time dependence factor $e^{-i \omega t}$ has been omitted. In the Regge-Wheeler gauge, the metric quantities $h^{(e)}_{0 \ell m}(r), h^{(e)}_{1 \ell m}(r), G_{\ell m}(r), h_{2 \ell m}(r)$ are set to be zero. The remaining metric quantities can be reconstructed by the odd and even parity master variables $\Psi^{\rm o}, \Psi^{\rm e}$, which are solutions of the master wave equations

\begin{align}
\left [ \partial^2_{r_*} +\omega^2 -V^{\rm o}(r) \right ] \Psi^{\rm o} = S^{\rm o} (r)\,,
\end{align}
and
\begin{align}\label{eq:mee}
\left [ \partial^2_{r_*} +\omega^2 -V^{\rm e}(r) \right ] \Psi^{\rm e} = S^{\rm e} (r)\,,
\end{align}
with $d r_* = (1-2M/r)^{-1} dr$. Here the source terms $S^{\rm o}, S^{\rm e}$ are explicitly
given in \cite{2003PhRvD..67j4018J}, and the potential terms are
\begin{align}
 V^{\rm o}(r) =& \left ( 1- \frac{2 M}{r}\right ) \left ( \frac{2(\lambda_l+1)}{r^2} - \frac{6 M}{r^3}\right ) \,,\nonumber \\
 V^{\rm e}(r) =& \left ( 1- \frac{2 M}{r}\right ) \nonumber \\
& \times \frac{2\lambda_l^2(\lambda_l+1)r^3+6 \lambda_l^2 M r^2+18 \lambda_l M^2 r + 18 M^3}{r^3(r \lambda_l+3)^2}\,,
\end{align}
with $\lambda_l=(\ell-1)(\ell+2)/2$. For the source trajectory considered here, the $\ell=2, m=\pm 2$ piece of metric perturbation dominates. In addition, the odd-parity source terms are zero, such that the odd-parity metric perturbations are also zero. We numerically solve the even-parity Master equation by applying ingoing boundary condition at horizon and outgoing boundary condition at infinity.
The results for the trajectory described in Table \ref{tab:orbit} is shown in Fig.~\ref{fig:psi}.
%, where the curves for type I orbits are qualitative the same.
The metric quantities are directly reconstructed based on the solutions of $\Psi^{\rm e}$.

\subsection{Phase space}

\begin{figure}
	\includegraphics[width=8.cm]{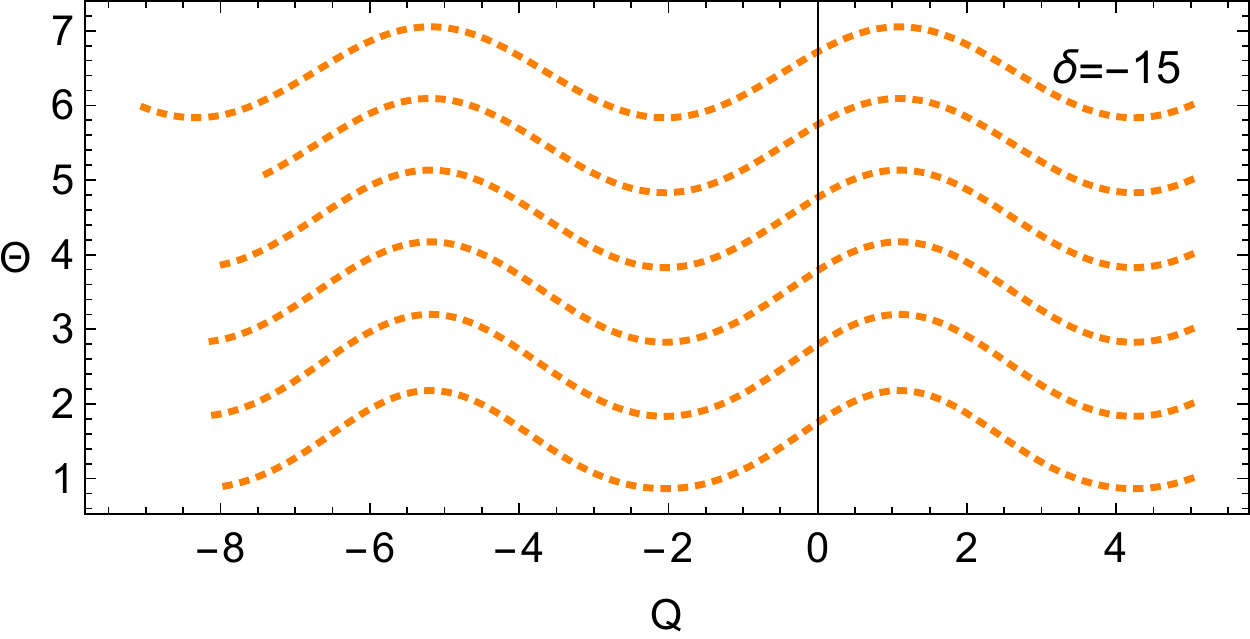}
	\includegraphics[width=8.cm]{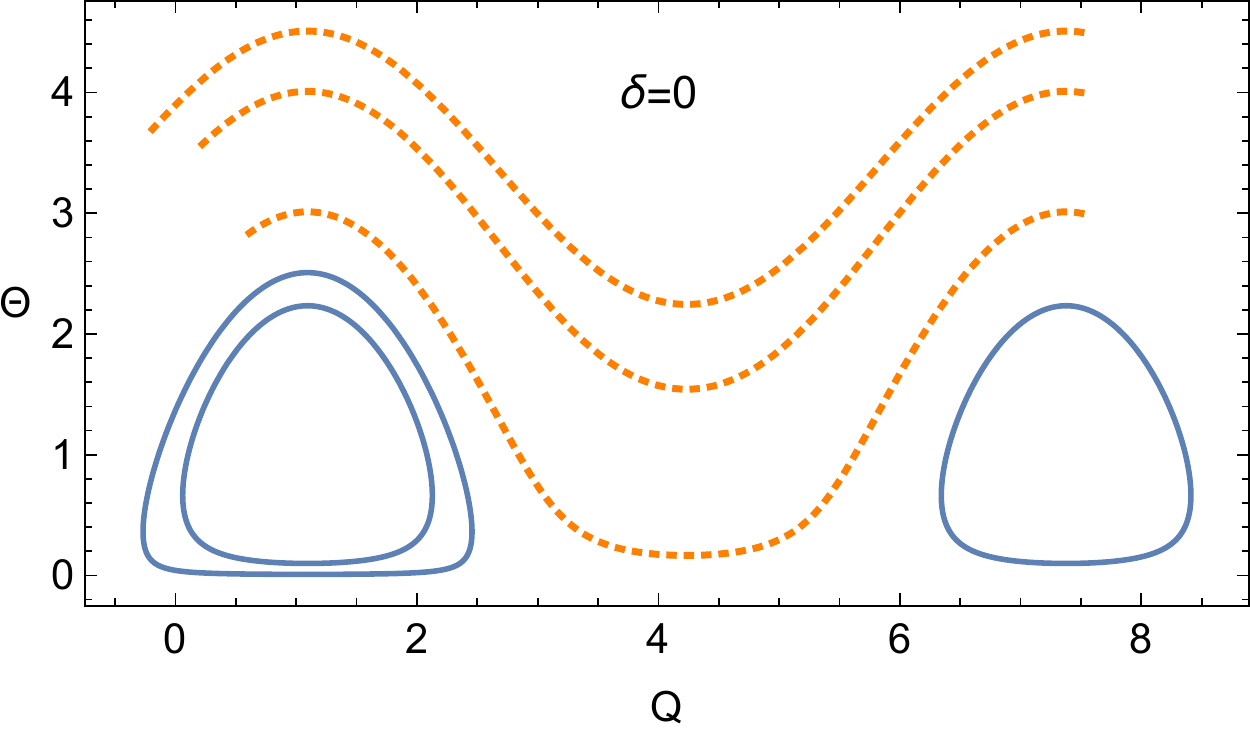}
	\includegraphics[width=8.cm]{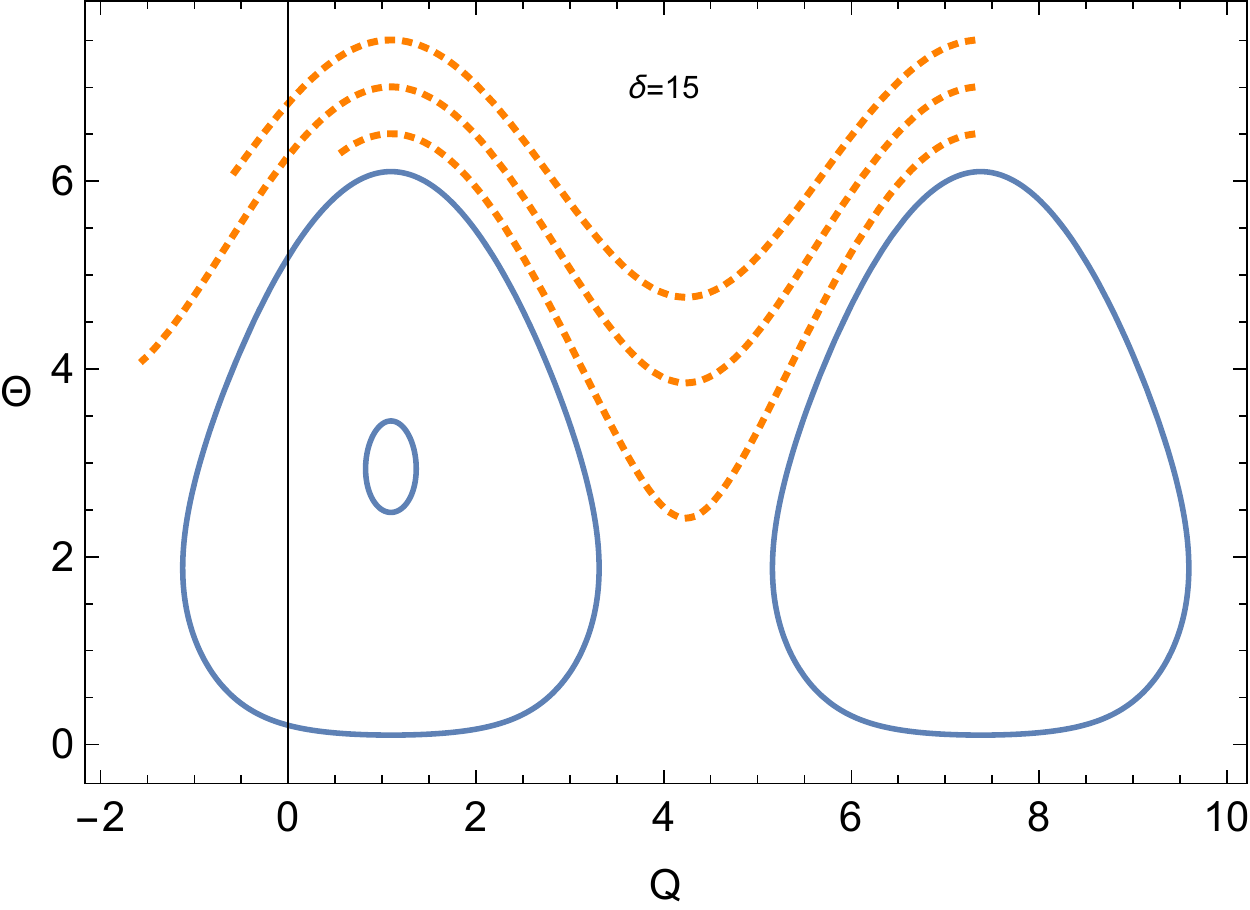}
	\caption{The trajectories in phase space corresponding to the effective Hamiltonian described by Eq.~\eqref{eq:hn}. For both $\delta =0$ and $\delta =15$ cases, there are two regimes in the phase space:  the libration regime (solid, blue lines)
		and  the rotation regime (dashed, orange lines). There is only a rotation regime for $\delta =-15$.}
	\label{fig:p1}
\end{figure}

With the trajectory in Table \ref{tab:orbit} and the reconstructed metric perturbations, we can explicitly write down the effective Hamiltonian.
For simplicity we assume that the masses of the resonant objects are the same $\mu=\mu'$, so that $\eta =\eta'$.
\begin{align}\label{eq:hn}
H_{\rm eff} \approx \alpha \Theta- 0.032 \eta \Theta^2+0.02\eta \sqrt{\Theta} (6\sin Q+3.1\cos Q)\,,
\end{align}
where this is an  approximate expression as we have only kept the dominant $\ell=2, m=\pm 2$ harmonics.
The conjugate momentum is given by $\Theta =\Gamma_r J_r/\mu$ in this particular example.
Next, we rescale the Hamiltonian by a factor $10^{-2} \eta$, which is equivalent to rescaling the time.
The new Hamiltonian is
\begin{align}
H'_{\rm eff} =  \delta \Theta- 3.2  \Theta^2+2 \sqrt{\Theta} (6\sin Q+3.1\cos Q)
\end{align}
%For illustration purpose, we define $\tilde{P} :=5\times 10^{-3} P$, with equation of motion
%\begin{align}\label{eq:eomr}
%\frac{d \tilde{P}}{d \hat{\lambda}} = 2 (-1.17 \cos Q+0.6 \sin Q)\,,
%\end{align}
with $\delta : = \alpha/(0.01\eta)$.
The phase space trajectories follow level curves of $H_{\rm eff}'$ and are shown in Fig. \ref{fig:p1} in terms of $(\Theta, Q)$.
The topology of the phase space is completely determined by $H_{\rm eff}'$ and depends only upon the value of $\delta$: There are cases where the phase space can be naturally divided into a ``rotation'' regime and a ``libration'' regime, and cases where there is only a rotation regime. The motion in the libration regime
is trapped, which corresponds to the mean motion resonance considered here. Different libration regimes are equivalent to each other due to the $Q \rightarrow Q+2 \pi$ symmetry of the effective Hamiltonian.

\begin{figure}
\includegraphics[width=7.5cm]{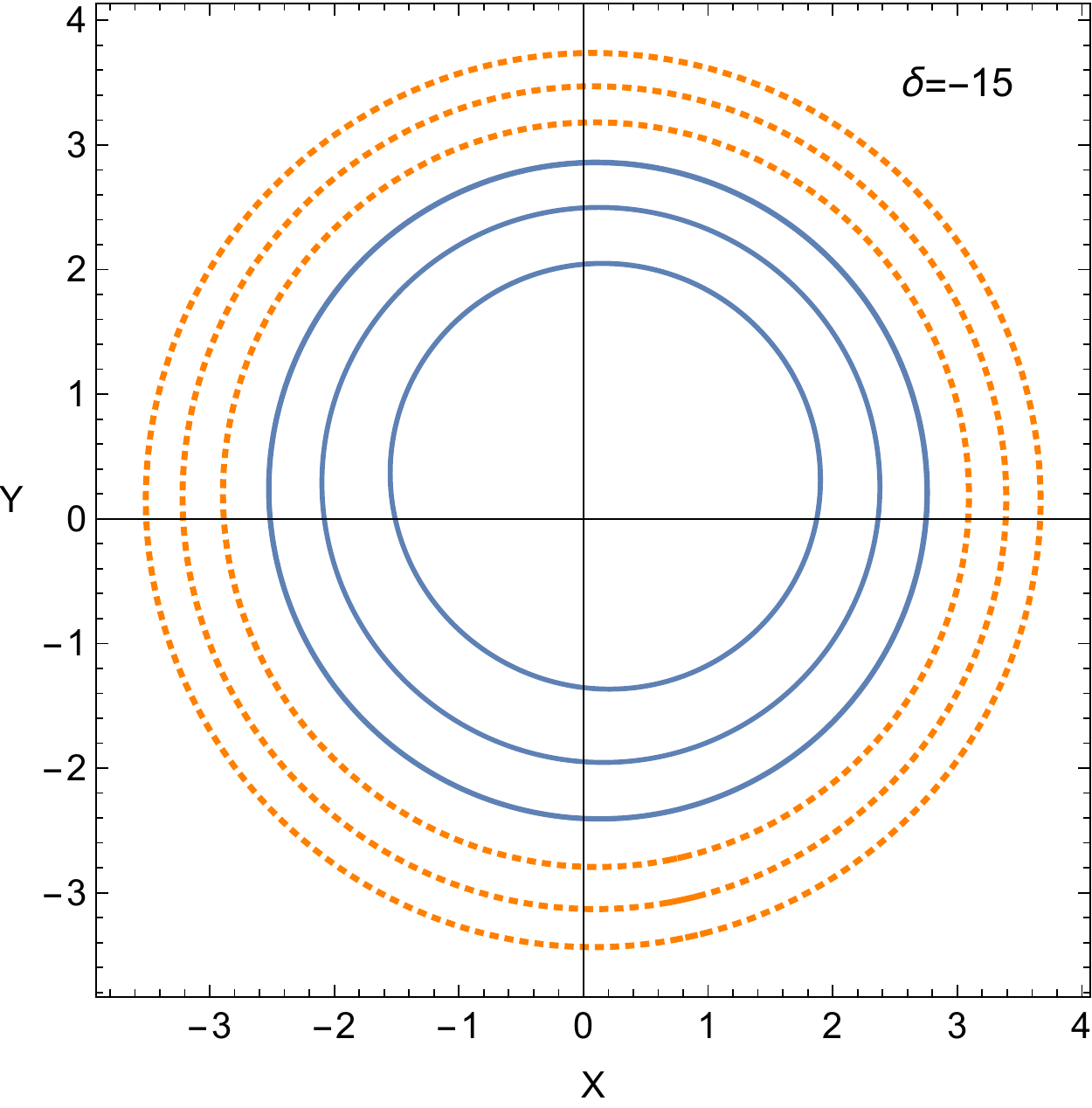}
\includegraphics[width=7.5cm]{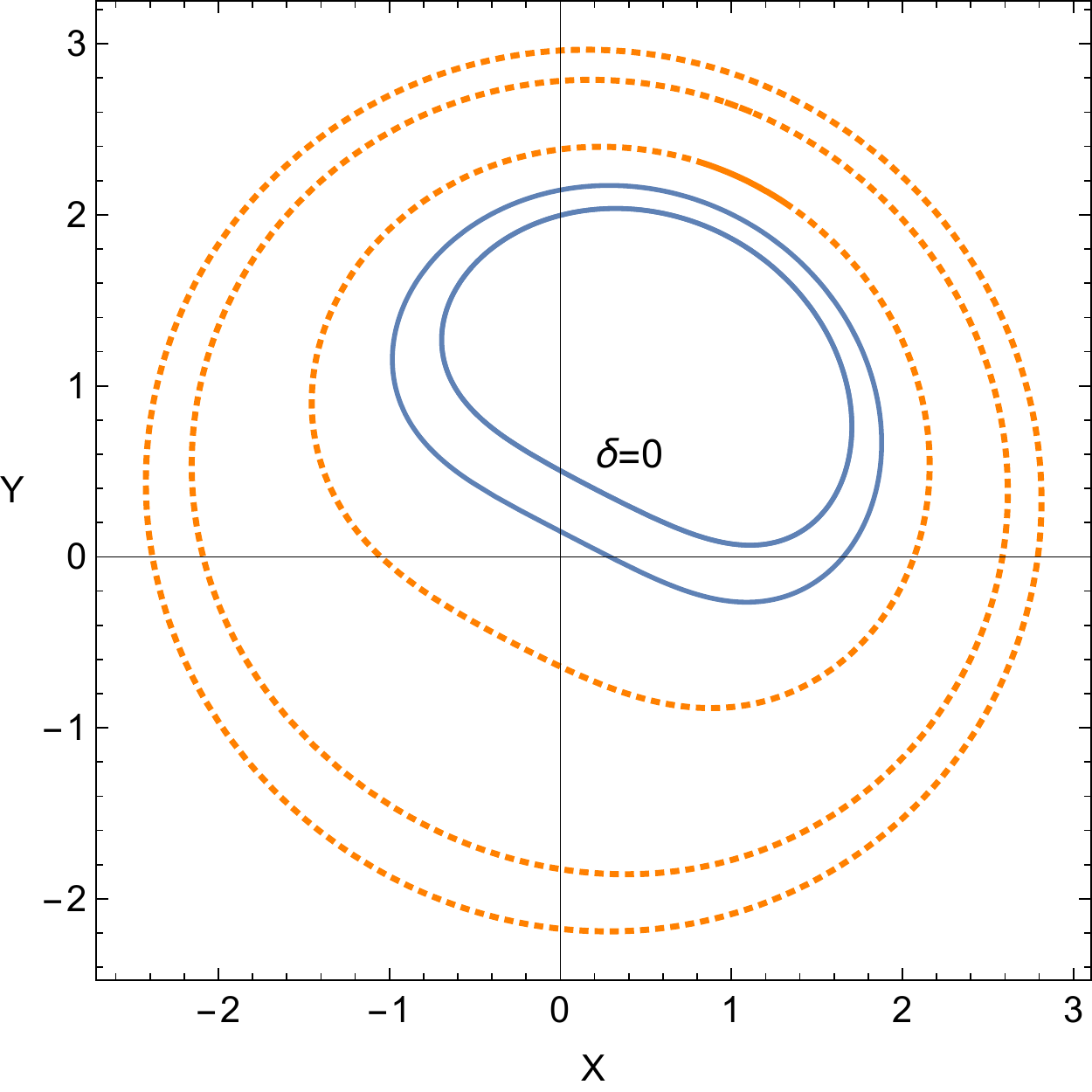}
\includegraphics[width=7.5cm]{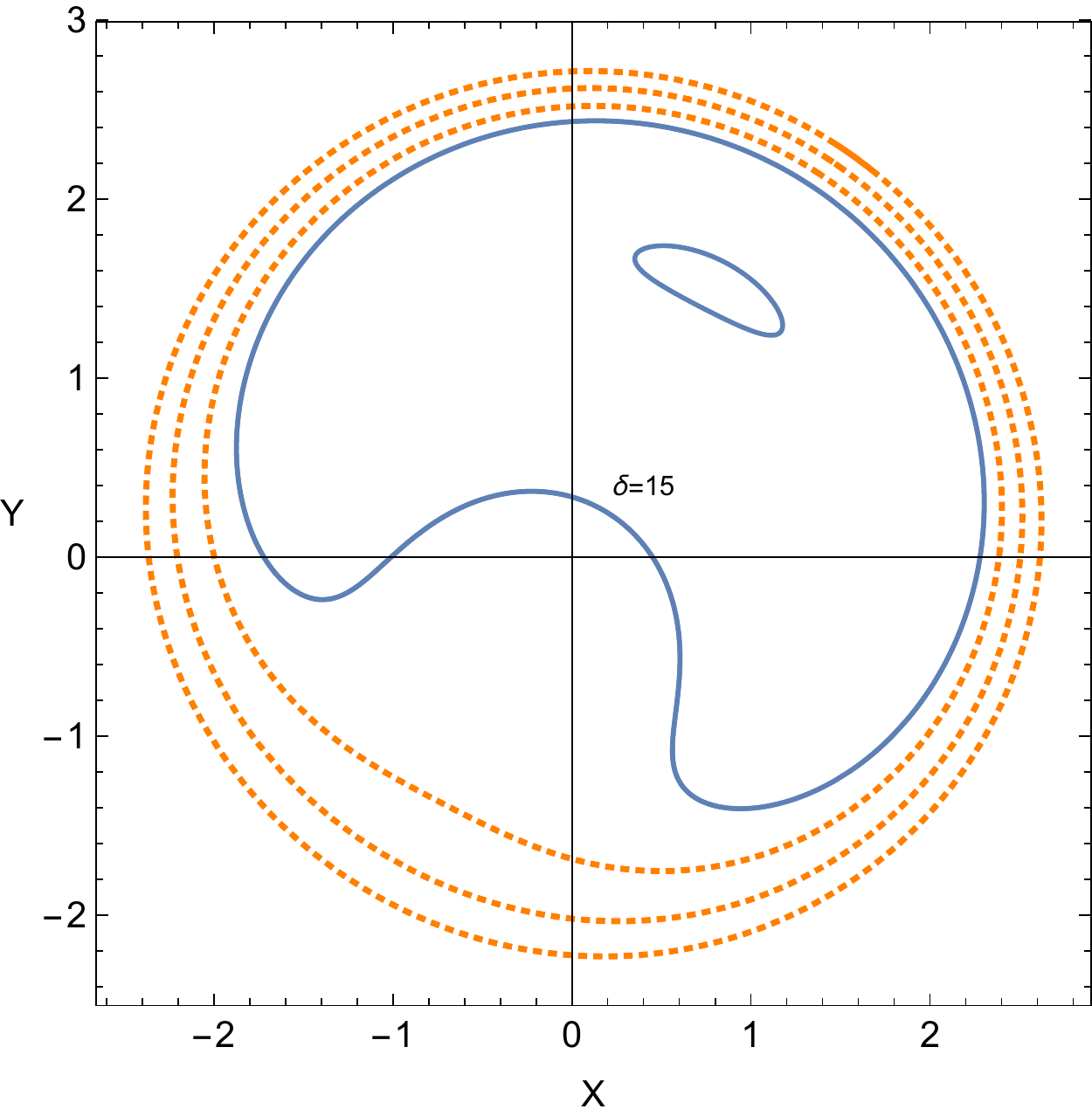}
\caption{Same phase space trajectories as in Fig.~\ref{fig:p1}, although the canonical variables are chosen as in Eq.~\eqref{eq:recan}.}
\label{fig:p2}
\end{figure}

For illustration purposes, we also  define
\begin{align}\label{eq:recan}
X = \sqrt{2\Theta} \cos Q, \quad Y = \sqrt{2\Theta} \sin Q\,,
\end{align}
with the corresponding phase space trajectories shown in Fig.~\ref{fig:p2}.
The origin in these plots corresponds to zero eccentricity, and the distance from the origin is proportional to the eccentricity.  The orbits in the rotation regime correspond to the cases that the resonance is broken.
%The tidal resonance considered in [] is a special case of the rotation trajectory.

%To investigate the effect of dissipation (radiation reaction, disk force, etc), and in light of the general Hamiltonian form in Eq.~\eqref{eq:hdis}, we modify the equation of motion in Eq.~\eqref{eq:eomr} by adding a constant term $\epsilon_{\rm dis}$:
%\begin{align}\label{eq:eomr}
%\frac{d \tilde{P}}{d \hat{\lambda}} = -\epsilon_{\rm dis}+2 (-1.17 \cos Q+0.6 \sin Q)\,,
%\end{align}
%and plot different phase space trajectories in Fig. \ref{fig:p2}. We find that adding the effect of dissipation decrease the size of ``rotation" regimes, which become disjoint to each other. This has important implication to the secular evolution of motion near the mean-motion resonance, as the area enclosed by a phase space trajectory is an adiabatic invariant. If the size of ``rotation" regime shrinks because of increasing dissipation, a trajectory that initially belongs to the ``rotation" regime may eventually cross the boundary and enter the ``libration" regime. Whenever this happens, the mean-motion resonance is broken.

In the effective Hamiltonian above we consider $\alpha$/$\delta$ as constant. However, to properly account for the secular dynamics, we also need to consider the parametric modification of $\alpha$/$\delta$ due to the secular change of the system's energy, angular momentum, etc.
When this is taken into account, the actual trajectories in phase space are not the closed trajectories in the $(X,Y)$ phase plane shown in Fig.~\ref{fig:p2}. Nonetheless, these level curves of the effective Hamiltonian are very useful as they serve as ``guiding'' trajectories for the evolution. In particular, from the trajectories it is clear that if one is near resonance so that the effective Hamiltonian describes the evolution of the system, but not on resonance yet, and one has $\delta<0$, a necessary (but not sufficient) condition for resonance to occur is that $\delta$ has to increase such that it becomes positive \cite{Murray-Dermott}. And vice versa, if $\delta$ is initially positive, $\dot{\delta}$ needs to be negative for resonance to occur.
In addition,  the action of an orbit, that is,
\begin{align}
	J = \oint \Theta \; d Q = \oint X \; dY \;,
\end{align}
is an adiabatic invariant of motion, and in this case is simply the area enclosed by a phase space trajectory in the $(X,Y)$ plane. The action is not conserved when the orbits evolves close to the resonant critical curve/separatrix (as there the period of the motion becomes infinite).
Therefore, for guiding trajectories which remain away from the critical curve, adiabatic changes in $\delta$ preserve the area enclosed by the trajectory  in the  $(X,Y)$-plane, even as its center moves.
After the critical curve has been crossed the action again becomes an approximate adiabatic invariant.
Based on these considerations, one can qualitatively predict the possible outcomes of capture into resonance.
For instance, consider conditions near resonance with $\delta<0, \dot{\delta}>0$ and with initial small eccentricity (in other words, the area enclosed by the guiding trajectory is small). As $\delta$ increases, the trajectory can stay in the circulation regime and `miss' the resonance or it can be captured into resonance. If capture occurs, the resulting eccentricity will be larger as the guiding trajectory will be off-center (see Fig.\ref{fig:p2}).
Hence, for the effective Hamiltonian $H_{\rm eff}'$, there is a significant change in the eccentricity due to capture into resonance.
In realistic situations, we also need to take into account the dissipative forces that drive the orbital migration, which likely affect the resonance capture and evolution as well. This is  seen in the numerical evolution of various resonances in Sec.~\ref{sec:numerics}. In that section, we also discuss in more detail the dynamics near resonance accounting for the evolution of the orbital parameters due to dissipation.

\section{Post-Newtonian Hamiltonian Formalism}
\label{sec:post-newtonian}

In this section, we derive an effective post-Newtonian Hamiltonian to analyze the  dynamics near mean motion resonance. We restrict ourselves to the case of two small bodies with masses $m_1$ and $m_2$ orbiting the central massive object $M$ in the equatorial plane. %In the extreme mass ratio limit we have $m_1, m_2 \ll M$.

The Hamiltonian of this system, which we denote by $H$ to distinguish it from the relativistic Hamiltonian $\mathcal{H}$, can be written as:
\begin{equation}
	H = H_1 + H_2 + H_{ {\rm int}}
\end{equation}
where $H_1$ is the Hamiltonian of body $m_1$, $H_2$ that of body $m_2$ and $H_{\rm int}$ is the interaction Hamiltonian. To make the different Newtonian orders explicit, we will (partially) reinstate factors of $c$ in this section, but the gravitational constant $G$ is still set to one. To first post-Newtonian order $H_1$ is given by \cite{Blanchet2006}
\begin{align}
	H_1 &= \frac{1}{2 m_1} \left(p_r^2 + \frac{p_\phi^2}{r^2}\right) - \frac{m_1 M}{r} \notag \\
	& \quad + \frac{1}{c^2} \left[- \frac{1}{8 m_1^3} \left(p_r^2 + \frac{p_\phi^2}{r^2}\right)^2 + \frac{m_1 M^2}{2 r} \right. \notag  \\
	&\left. \qquad \qquad \quad - \frac{3 M}{2 m_1 r} \left(p_r^2 + \frac{p_\phi^2}{r^2}\right)\right]  + \mathcal{O}\Big(\frac{1}{c^4}\Big) \label{eq:h1pn}
\end{align}
and similarly for the second body with the relevant quantities denoted with an underbar.
We will return to the explicit form of the interaction Hamiltonian, but first we will rewrite $H_1$ (and $H_2$) in terms of Poincar\'e variables. These variables are a linear combination of the standard action-angle variables associated to the coordinates $r, \phi$ and  have been extremely valuable in the study of planetary dynamics.\footnote{Delauney variables --- another frequently used set of variables in celestial mechanics --- are yet another linear combination of the standard action-angle variables. However, these have the disadvantage that they are not well-defined for orbits with vanishing eccentricities.}
In order to perform the transformation to Poincar\'e variables, we first observe that $\phi$ is a cyclic coordinate so that $p_\phi$ is constant and make a canonical transformation to action-angle variables
\begin{align}
  J_r &: = \frac{1}{2 \pi} \oint p_r \; dr  \label{eq:defJr}\\
  J_\phi &: = \frac{1}{2 \pi} \oint p_\phi \; d\phi = p_\phi \; .
\end{align}
Using the fact that $H_1$ is conserved and denoting this constant $H_1$ by $E$ (with $E<0$ as we consider bound orbits), we write $p_r = p_r (r,E,p_\phi)$
\begin{align}
p_r & = \pm \frac{\sqrt{-2 m_1 E}}{r} \sqrt{(r-\rmin)(-r+\rmax)} \label{eq:pr} \\
& \quad \pm \frac{1}{c^2} \frac{r}{2 \sqrt{-2 m_1 E}} \frac{ E^2 + \frac{8 m_1 M E}{r} + \frac{6 m_1^2 M^2}{r2}}{\sqrt{(r-\rmin)(-r+\rmax)}}  + \mathcal{O}\Big(\frac{1}{c^4}\Big) \notag
\end{align}
with $\rmin$ and $\rmax$ defined by the requirement that the Newtonian part of $p_r$ vanishes
\begin{align}
 r_{\pm}= \frac{m_1 M  \pm \sqrt{m_1^2 M^2 + \frac{2 p_\phi^2 E}{m_1}}}{-2 E} .
\end{align}
Substituting Eq.~\eqref{eq:pr} into the definition for $J_r$, and performing the relevant integrals (using standard contour integration), we obtain:
\begin{align}
J_r &= - p_\phi + \frac{m_1^2 M}{\sqrt{- 2 m_1 E}} \\
& \quad + \frac{1}{c^2} \left[ - \frac{15}{8} M \sqrt{- 2 m_1 E} + 3 \frac{m_1^2 M^2}{p_\phi}\right] + \mathcal{O}\Big(\frac{1}{c^4}\Big) \; . \notag
\end{align}
In principle, we should have taking into account that $r_\pm$ is shifted by post-Newtonian corrections and therefore that the integration limits in Eq.~\eqref{eq:defJr} are also shifted and not simply $r_\pm$. A careful analysis of these `edge' contributions shows that their change is sub-dominant and we shall neglect these corrections.
From the standard action-angle variables $\{ q_r,q_\phi, J_r, J_\phi \}$, we perform a canonical transformation to the Poincar\'e variables
\begin{equation}
\begin{split}
& \gamma = q_r - q_\phi \\
& \lambda = q_\phi
\end{split}
\qquad \qquad
\begin{split}
&\Gamma = J_r \\
& \Lambda = J_r + J_\phi \;.
\end{split}
\end{equation}
The generating function of this transformation is
\begin{align}
  F  = (q_r - q_\phi) \Gamma + q_\phi \Lambda \; .
\end{align}
After these canonical transformations, the Hamiltonian for the body with mass $m_1$ is
\begin{align}
 H_1 = - \frac{m_1^3 M^2}{2 \Lambda^2} + \frac{1}{c^2} \frac{3 m_1^5 M^4}{8 \Lambda^4} \frac{5 \Gamma + 3 \Lambda}{\Gamma - \Lambda}  + \mathcal{O}\Big(\frac{1}{c^4}\Big) \; .
\end{align}
This Hamiltonian recovers the well-known post-Newtonian precession rate to first post-Newtonian order \cite{Blanchet2006}:
\begin{align}
\dot{\gamma} & =  \frac{\partial H_1}{\partial \Gamma} = - \frac{3}{c^2} \frac{m_1^5 M^4}{\Lambda^3 (\Gamma-\Lambda)^2} + \mathcal{O}\Big(\frac{1}{c^4}\Big) = \omega_r - \omega_\phi \notag \\
\dot{\lambda} & =  \frac{\partial H_1}{\partial \Lambda} = \frac{m_1^3 M^2}{\Lambda^3} \notag  \\
&\qquad \qquad - \frac{1}{c^2} \frac{3 m_1^5 M^4}{2 \Lambda^5} \frac{5 \Gamma^2 - 4 \Lambda \Gamma-3 \Lambda^2}{ (\Gamma-\Lambda)^2} + \mathcal{O}\Big(\frac{1}{c^4}\Big)  = \omega_\phi \notag
\end{align}
where
\begin{align}
\omega_r  & = \frac{(-2E)^{3/2}}{m_1^{3/2}M} \left(1 + \frac{1}{c^2} \frac{15 E}{4 m_1}\right) +  \mathcal{O}\Big(\frac{1}{c^4}\Big) \label{eq:omegar} \\	
\omega_\phi  & = \frac{(-2E)^{3/2}}{m_1^{3/2}M} \left(1 + \frac{1}{c^2} \left(\frac{15 E}{4 m_1} + 3 \frac{m_1^2 M^2}{p_\phi^2}\right) \right) +  \mathcal{O}\Big(\frac{1}{c^4}\Big) \label{eq:omegaphi}
\end{align}
(see Eq. (345) in \cite{Blanchet2006} after identifying $\omega_r$ with $n$ and $\omega_{\phi}$ with $n K$).

While this formalism allows us to investigate resonance behavior in a variety of circumstances, here we focus on a particular simple case: a first order exterior $j-1:j$ resonance, whereby the internal object $m_1$ completes $j-1$ cycles and the external object $m_2$ completes $j$ cycles before the system returns to its original state. We focus on first order resonances as lower order mean motion resonances are more important than higher order ones for the orbital dynamics of three-body systems.
%, as the amplitude of higher order resonances is much smaller.
The interaction Hamiltonian describes the gravitational interaction between the body with $m_1$ and that with $m_2$. It contains terms that can be classified as short period, secular and resonant. The short period terms vanish after orbit averaging and contribute negligible to the long term dynamics of the system. Therefore, for most purposes these short term terms can be ignored.
The secular and resonant terms are both important for a complete understanding of the orbital dynamics. However, to understand the dynamics of the system near resonance,  the resonant terms dominate so that we can consider the following simple  form for the interaction Hamiltonian \cite{Murray-Dermott}
\begin{align}
 H_{\rm int} &=  - f_d \frac{m_1 m_2^3 M}{\underline{\Lambda}^2} \sqrt{\frac{2 \underline{\Gamma}}{\underline{\Lambda}}} \cos \left((1-j) \underline{\lambda} + j \lambda + \underline{\gamma} \right) \notag \\
 &\qquad  +\mathcal{O}\Big(\frac{1}{c^2}\Big) \; ,
\end{align}
with $f_d$ indicating the strength of the interaction. Although $f_d$ is in principle a function of the ratio of the semi-major axes of the two orbiting bodies, its functional dependence is not relevant to the orbital dynamics and we will treat it as a constant \cite{peale1976}.

The expressions so far are valid for any value of eccentricity. To further simplify the analysis, we expand the above Hamiltonian for small eccentricities following similar steps in \cite{Murray-Dermott}.
In order to perform the small eccentricity expansions, we yet again make a canonical transformation:
\begin{equation}
\begin{split}
&\theta_1 = (1-j) \underline{\lambda} + j \lambda + \gamma \\
&\theta_2 = (1-j) \underline{\lambda} + j \lambda + \underline{\gamma} \\
&\theta_3 = \lambda \\
&\theta_4 = \underline{\lambda}
\end{split}
\qquad \quad
\begin{split}
& \Theta_1 = \Gamma \\
& \Theta_2 = \underline{\Gamma} \\
& \Theta_3 = \Lambda -j (\Gamma + \underline{\Gamma})  \\
& \Theta_4  = \underline{\Lambda}- (1-j) (\Gamma + \underline{\Gamma}) \; .
\end{split}
\end{equation}
For an exterior resonance, the new coordinates $\theta_1, \theta_3$ and $\theta_4$ are all cyclic so that the associated momenta $\Theta_1, \Theta_3$ and $\Theta_4$ are constant. Therefore, the problem has effectively been reduced to a system with one degree of freedom described by $\{\theta_2, \Theta_2\}$. (For interior resonances, the story is very similar and the effective degree of freedom is described by $\{\theta_1, \Theta_1\}$.) By relating the momenta to the orbital elements, and in particular to the Newtonian eccentricity $e$, we find that $\Theta_1, \Theta_2 \sim \mathcal{O}(e^2)$ whereas $\Theta_3, \Theta_4 \sim \mathcal{O}(1)$. Therefore, given that $\Theta_1$ is constant and small, we will neglect $\Theta_1$ as this has little effect on the dynamics.
Expanding the Hamiltonian to second order in $\frac{\Theta_2}{\Theta_3}$ and $\frac{\Theta_2}{\Theta_4}$, and performing a (partial) transformation back to the Poincar\'e variables using $\Theta_3 \approx \Lambda$ and $\Theta_4 \approx \underline{\Lambda}$, we find that the Hamiltonian describing near resonance behavior is:
\begin{align}
	\label{eq:effective-hamiltonian-pn}
 H = \alpha  \underline{\Gamma} + \beta  \underline{\Gamma}^2 + \kappa \sqrt{2  \underline{\Gamma}} \cos \theta_2 + \mathcal{O}\Big(\frac{1}{c^4}, f_d^2, e^3 \Big)
\end{align}
\begin{align}
\alpha &:=j \frac{m_1^3 M^2}{\Lambda^3} + (1-j) \frac{m_2^3 M^2}{\underline{\Lambda}^3} \notag \\
&\qquad + \frac{1}{c^2} \left[\frac{9}{2} j \frac{m_1^5 M^4}{\Lambda^5} + \frac{3}{2} (1-3j) \frac{m_2^5 M^4}{\underline{\Lambda}^5} \right] \notag \\
\beta &:= - \frac{3j^2}{2} \frac{m_1^3 M^2}{\Lambda^4} - \frac{3(1-j)^2}{2}  \frac{m_2^3 M^2}{\underline{\Lambda}^4} \notag \\
&\qquad + \frac{1}{c^2} \left[-\frac{45}{4}j^2 \frac{m_1^5 M^4}{\Lambda^6} \right. \notag \\
&\qquad \qquad  \quad \left. + \frac{3}{4} (1+ 10j - 15 j^2) \frac{m_2^5 M^4}{\underline{\Lambda}^6} \right] \notag  \\
\kappa & := - f_d \frac{m_1 m_2^3 M}{ \underline{\Lambda}^{5/2}} \; .  \label{eq:coeff-hamiltonian}
\end{align}
The constant $\alpha$ measures the proximity to resonance as resonance occurs when the time-derivative of the resonant argument $\theta_2$ vanishes.
In the Newtonian limit, it is clear that $\alpha$ measures the proximity to resonance after noting that $\Lambda = m_1 \sqrt{M a}$ and $\Gamma= m_1 \sqrt{M a} (1-\sqrt{1-e^2})$ with $a$ the semi-major axis and $e$ the eccentricity of the unperturbed orbit of the object with mass $m_1$ around the central massive object $M$, so that $\alpha_{\rm Newton} = (1-j) \; \omega + j \; \underline{\omega}$.
In fact, due to the gravitational interaction between the two bodies the radial and azimuthal frequencies are not degenerate. Similarly, post-Newtonian corrections also break the degeneracy between the radial and azimuthal frequencies. Therefore, we expect that at post-Newtonian order $\alpha$ can be written as:
\begin{align}
\alpha & = \underline{\omega}_r - j \underline{\omega}_{\phi} + j \omega_{\phi} \label{eq:alphaintermsofomegas}
\end{align}
To show this is indeed the case, we need to relate the expression for $\alpha$ in terms of the Poincar\'e variables to the orbital frequencies in the small eccentricity limit (as the Hamiltonian is also derived in the small eccentricity limit).
We do this by writing both $\Lambda,\underline{\Lambda}$ and the orbital frequencies in terms of the gauge-invariant energy of the orbits $E, \underline{E}$.
First, we note that at leading Newtonian order,  $\Gamma \sim \mathcal{O}(e^2)$ so that $p_\phi = \Lambda - \Gamma = \Lambda + \mathcal{O}(e^2)$. Therefore, in the small eccentricity limit we can express  $\Lambda$ entirely in terms of $E$ (and similarly $\underline{\Lambda}$)
\begin{align}
%\Gamma & = - p_\phi + \frac{m_1^{3/2}M}{\sqrt{-2 E}} +  \frac{1}{c^2}  \frac{9}{8} M \sqrt{- 2 m_1 E} + \mathcal{O}(\frac{1}{c^{4}},e^2)  \\
\Lambda & =  \frac{m_1^{3/2}M}{\sqrt{-2 E}} + \frac{1}{c^2}  \frac{9}{8} M \sqrt{- 2 m_1 E} + \mathcal{O}(\frac{1}{c^{4}},e^2)  \label{eq:lambda-small-e}
\end{align}
where we replaced $p_\phi$ in the post-Newtonian part by its expression in the circular limit, that is, $\Lambda = m_1^{3/2} M/\sqrt{-2E} + \mathcal{O}(c^{-2})$.
The orbital frequencies can also be expressed entirely in terms of $E$. In the small eccentricity limit the relation between $\omega_r$ and $E$ in Eq.\eqref{eq:omegar} does not change, but $\omega_{\phi}$ simplifies:
\begin{align}
\omega_\phi  & = \frac{(-2E)^{3/2}}{m_1^{3/2}M} \left(1 - \frac{1}{c^2} \frac{9 E}{4 m_1}\right) +  \mathcal{O}\Big(\frac{1}{c^4}, e^2\Big) \; .
\end{align}
Writing $\alpha$ in Eq.~\eqref{eq:coeff-hamiltonian} in terms of $E$ and using the  expressions for $\omega_r$ and $\omega_{\phi}$ in terms of $E$, we find that indeed
\begin{align}
\alpha - (\underline{\omega}_r - j \underline{\omega}_{\phi} + j \omega_{\phi}) = \mathcal{O}(c^{-4}, e^2).
\end{align}
This establishes that the expectation in Eq.~\eqref{eq:alphaintermsofomegas} is correct. It agrees with its fully relativistic counterpart on a Schwarzschild spacetime.

Higher order resonances of order $N$, that is, of the form $j-N:j$ slightly alter the numerical value of the coefficients $\alpha$ and $\beta$ and change the power of
$2 \Gamma$ in the interaction Hamiltonian to $(2\Gamma)^{N/2}$. Since $\Gamma \sim \mathcal{O}(e^2)$, this demonstrates why the orbital dynamics are dominated by lower order resonances.

Interior resonances can be treated very similarly. The resulting Hamiltonian will have the same form as in Eq.~\eqref{eq:effective-hamiltonian-pn}, but the constants will be slightly different.
%They results in regression of the pericentre of the small mass on the exterior orbit.

\section{Resonance Capture, Evolution and Escape }
\label{sec:numerics}

The capture, evolution and escape of mean motion resonance have been extensively discussed in planetary systems.
The capture only happens if it is a converging migration, in which case the ratio between the semi-major axes passes through the resonance value
towards one \cite{Murray-Dermott}. The migration could be driven by tidal interaction between the planets with the host star, or planets with the proto-planetary disk. The capture is easier if the initial eccentricities of the planets are small, although there are studies showing that large-eccentricity captures are still possible \cite{mmrhighe2017}. On the other hand, it has been shown that even when the eccentricities are very small, resonance capture may fail if the migration speed is too fast \cite{friedland2001migration}.

After the resonance capture, the locked pair of objects may migrate together within a disk. Depending on the dissipation mechanism, e.g. the disk force, and the system parameters, the resonance is sustained or breaks. The duration of resonances is related to the puzzle that most planets in multi-planet systems observed by {\it Kepler} spacecraft do not reside in mean motion resonances \cite{fabrycky2014architecture,goldreich2014overstable}.
The analysis in \cite{goldreich2014overstable} shows that under disk-planet interaction with characteristic semi-major axis damping rate $1/\tau_a$ and eccentricity damping rate $1/\tau_e$,  an exterior $j-1:j$ resonance is permanently sustained if
\begin{align}\label{eq:con1}
\eta' > \frac{j-1}{\sqrt{3} j^{3/2} c} \left( \frac{\tau_e}{\tau_a}\right )^{3/2}
\end{align}
with $c\approx 0.8 j$ and $\eta'$ the mass ratio of the outer object and the central massive object. On the other hand, if
\begin{align}\label{eq:con2}
\eta' < \frac{(j-1)^2}{8\sqrt{3} j^{3/2} \beta} \left( \frac{\tau_e}{\tau_a}\right )^{3/2}
\end{align}
 the resonance is only sustained for a duration proportional to the eccentricity damping timescale $\tau_e$. If the mass ratio $\eta'$ resides between the thresholds in Eq.~\eqref{eq:con1} and Eq.~\eqref{eq:con2}, the resonance is permanently sustained with a finite libration amplitude in the phase space.

Applying the insights from planetary dynamics to stellar-mass black holes near massive black holes, we immediately observe that the condition for sustained resonance in Eq.~\ref{eq:con1} is difficult to achieve with purely gravitational radiation damping, in which case  $\tau_e$ is comparable to $\tau_a$ unless the orbit is highly eccentric.
Therefore, the astrophysical environment of such systems is critical for sustained resonance to occur.
Stellar-mass black holes in galactic nuclei may migrate towards the central massive black hole due to mass segregation effects, dynamical friction and/or interaction with a possible accretion disk around the massive black hole.
%The gravitational interaction between these stellar-mass black holes, with the help of gravitational-wave damping and/or disk interaction, may also induce resonance locking between these objects.
Here we will consider a scenario with a thin-disk profile around the massive black hole, and two stellar-mass black hole (SBH) orbiting within the disk. In particular, following the description in \cite{kocsis2011observable}, we consider two types of thin-disk model: $\alpha$-disks and $\beta$-disks.

In the $\alpha$-disk model, the viscous stress is parameterized as $t_{t\phi} =-(3/2) \alpha\, p_{\rm tot}$ with $t_{t\phi}$ the viscous shear stress in the azimuthal direction,  $\alpha$ a dimensionless constant and $p_{\rm tot}$ the total pressure. The surface density $\Sigma$ of the $\alpha$-disk is
\begin{align}
\Sigma \sim 5.9 \times 10^{-21} M_{\odot}^{-1} \alpha_1^{-1} \dot{m}_{\bullet 1}^{-1} \bar{r}_{10}^{3/2} \,,
\end{align}
and the disk scale height $H$ is
\begin{align}\label{eq:scaleh}
H \sim 1.5 \times 10^{5} M_{\odot} \dot{m}_{\bullet 1} M_{\bullet 5} \,,
\end{align}
where we have defined $\alpha_1 : = \alpha/(0.1)$, $\bar{r}_{10} : = r/(10M$), $M_{\bullet 5} := M/(10^5 M_\odot)$,  and $\dot{m}_{\bullet 1}:=\dot{M}/(0.1 \dot{M}_{\bullet {\rm Edd}})$, with $\dot{M}_{\bullet {\rm Edd}}$ being the Eddington accretion rate.
The main difference between $\alpha$- and $\beta$-disks is the description of their viscous stress. For $\beta$-disks, the viscous stress is assumed to be $t_{t\phi} =-(3/2) \alpha\, p_{\rm gas}$, so that only the gas pressure $p_{\rm gas}$ contributes to the viscous stress instead of the total pressure $p_{\rm tot}$. As a result, the disk surface density is now given by
\begin{align}
\Sigma \sim 1.4 \times 10^{-17} M_{\odot}^{-1} \alpha_1^{-4/5} \dot{m}_{\bullet 1}^{3/5} \dot{M}_{\bullet 1}^{1/5} \bar{r}_{10}^{3/5} \,,
\end{align}
while the disk scale height is the same as Eq.~\eqref{eq:scaleh}.

There are two main types of disk-SBH interactions. The first is the accretion-induced force, where the Bondi accretion into the SBH brings in additional momentum and energy. The second force is known as Type I ``migration force'' and comes from the gravitational interaction between the SBH and the induced density waves in the disk (see discussions on Lindblad and co-rotational resonance in \cite{GT1979,GT1980}). Both of them predict that (with different $C$)
\begin{align}
\frac{1}{\tau_a} &=\frac{1}{\omega} \frac{d \omega}{dt} \sim C \;  \eta \; \eta_d \left( \frac{a}{H} \right )^2 \omega \,,\nonumber \\
\frac{1}{\tau_e} &=\frac{1}{e} \frac{d e}{dt} \sim C \; \eta \; \eta_d \left( \frac{a}{H} \right )^4 \omega \,,
\end{align}
with $a$ being the semi-major axis, $\eta_d =\Sigma a^2/M$  the disk to central black hole mass ratio, and the constants $C$ are $\mathcal{O}(1) -\mathcal{O}(10)$.
As the scale height $H$ in thin-disk models is constant, $\tau_e$ can be much smaller than $\tau_a$ for wide orbits ($a \gg H$), so that Eq.~\eqref{eq:con1} is satisfied and mean motion resonance is sustained.
For a central black hole with mass $M \sim 10^6 M_\odot$ and accretion rate $\dot{m}_{\bullet 1} \sim 1$, the gravitational radiation reaction becomes dominant for $r \le 100M $ for $\alpha$-disks and $r \le 30M$ for $\beta$-disks.
For radii larger than the critical radius the disk force is more important and sustained locking of the mean motion resonance becomes possible.

\begin{figure*}
	\centering
\subfloat{
	\includegraphics[width=8.5cm]{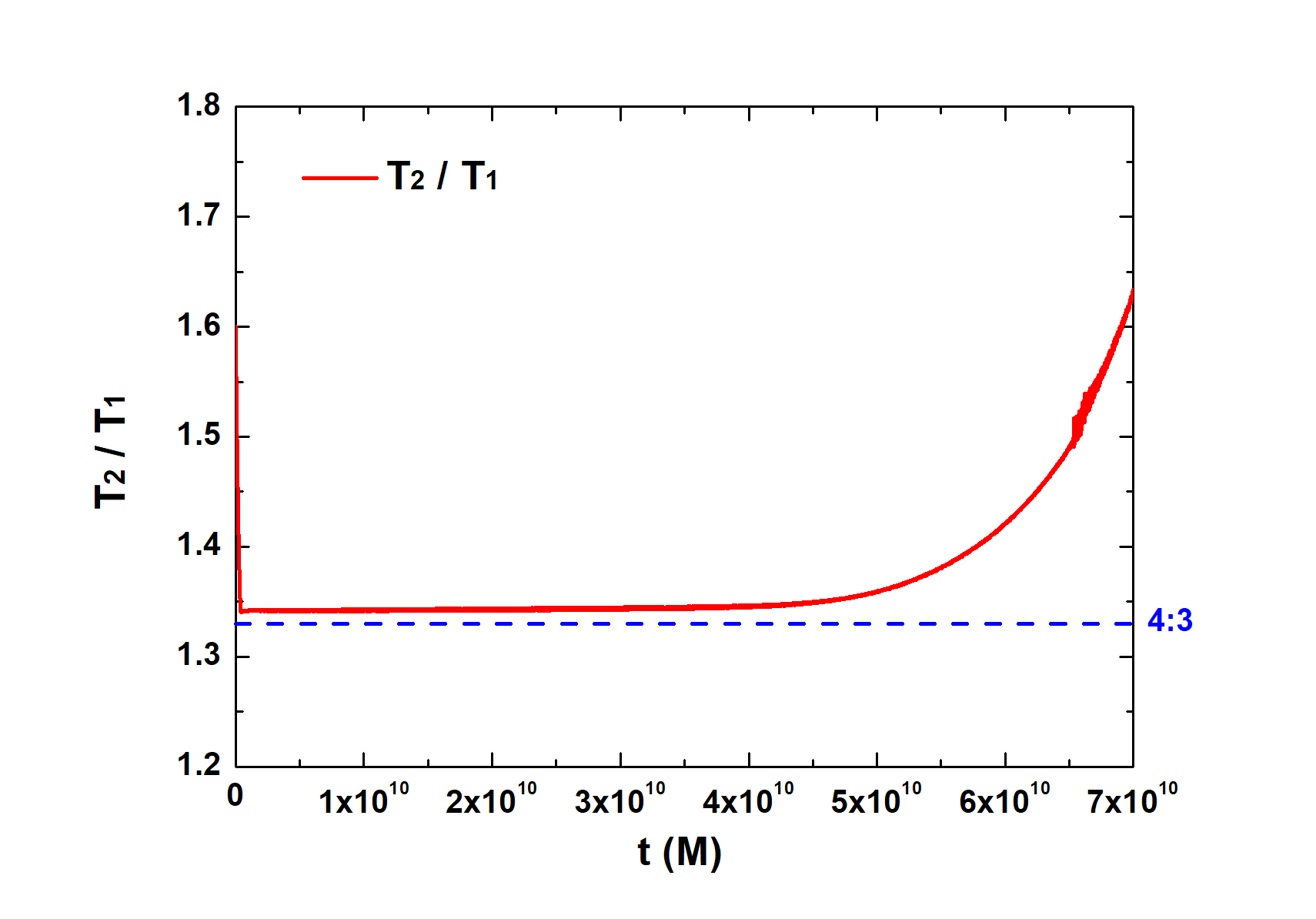}
}
\subfloat{
	\includegraphics[width=8.5cm]{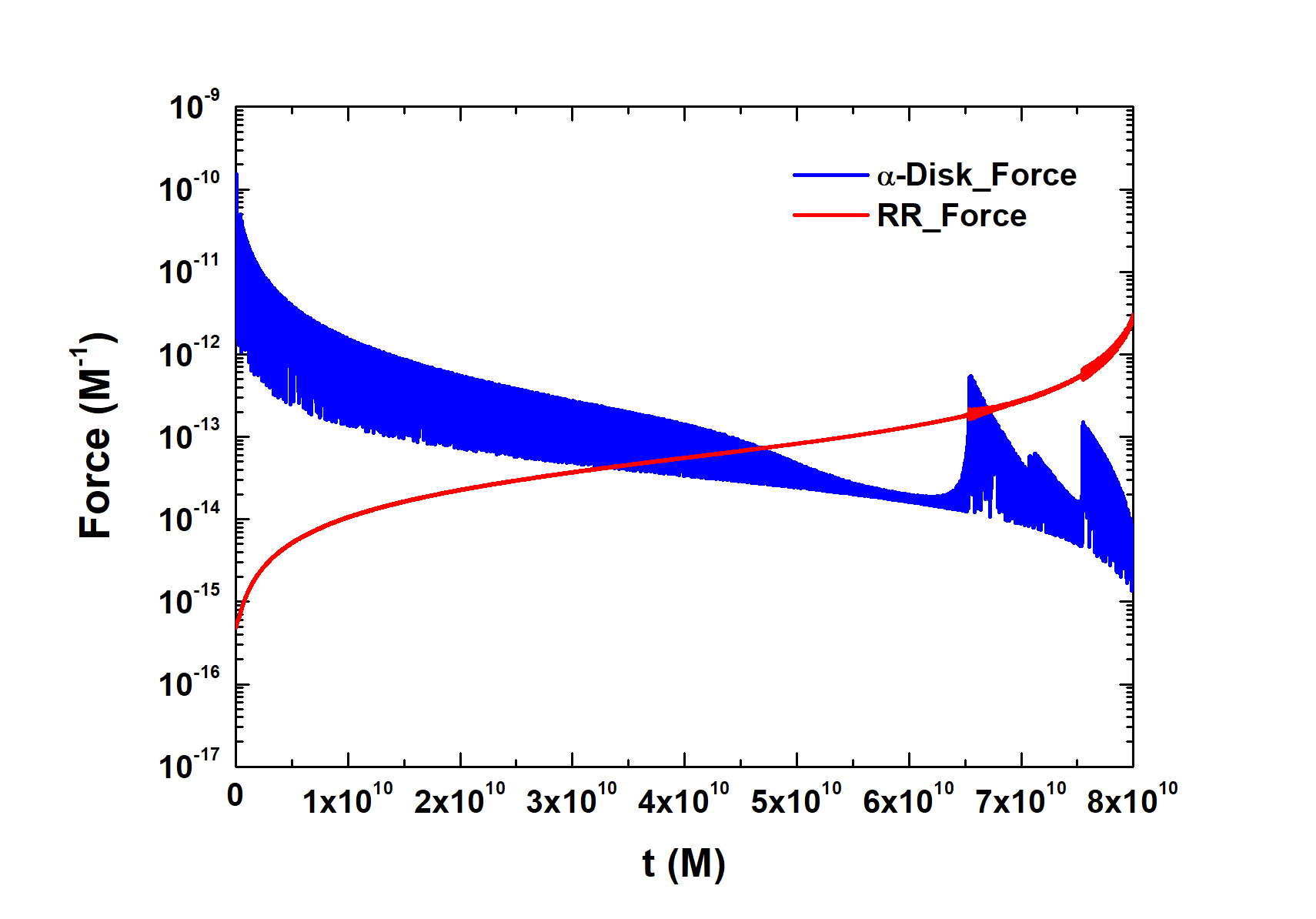}
}
\\
\subfloat{
	\includegraphics[width=8.5cm]{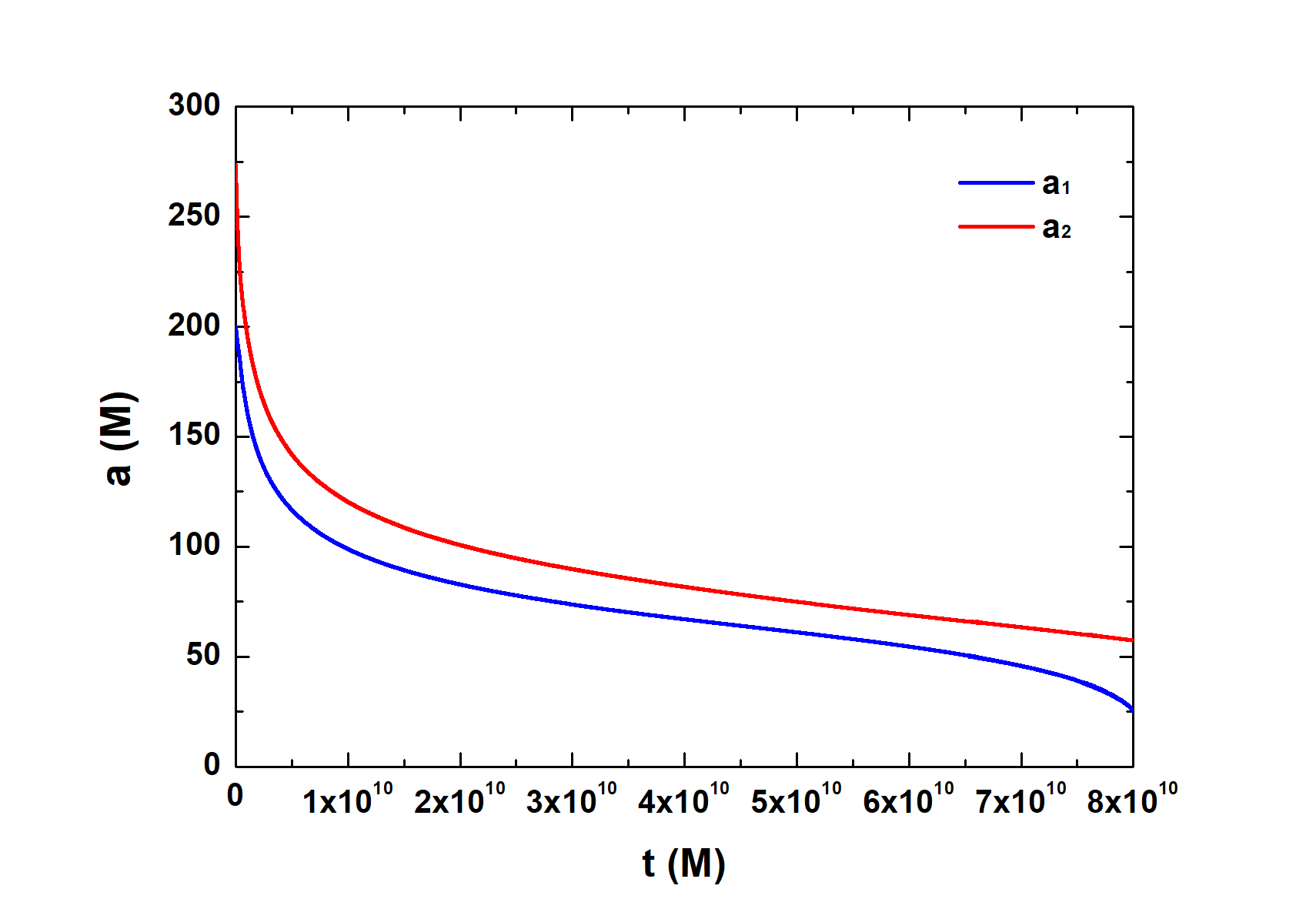}
}
\subfloat{
	\includegraphics[width=8.5cm]{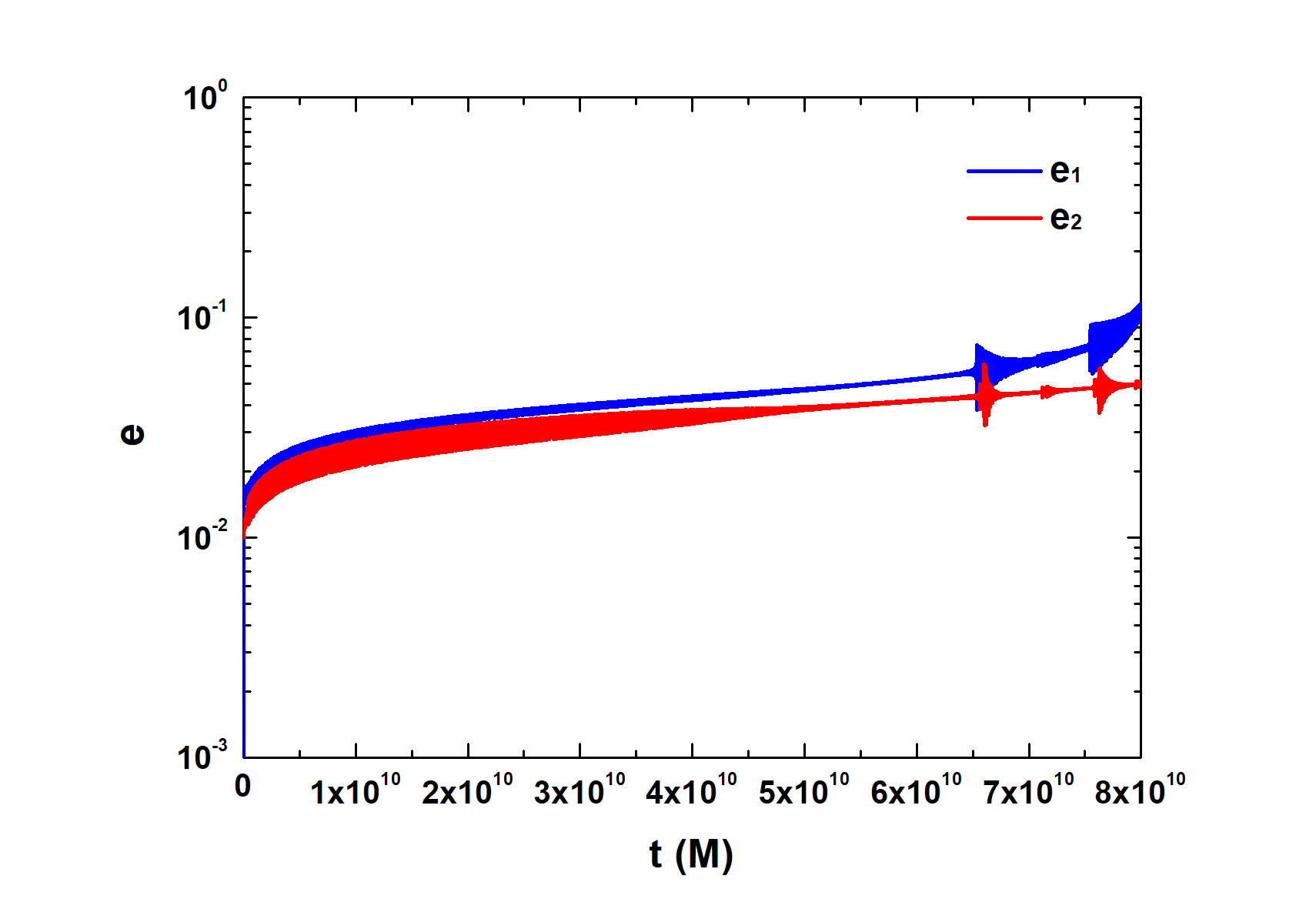}
}
\caption{Orbital evolution with the initial radius $a$  of the inner object equal to $200 M$, the semi-major axis $\underline{a}$ of the outer object equal to $273.5828M$ and its eccentricity $\underline{e}$ equal to 0.01.
The disk profile is modeled by an $\alpha$-disk.
Top left panel: the ratio between the periods of the two SBHs as a function of time showing that the system is captured into $3:4$ resonance.
Top right panel: the magnitude of the disk force and the gravitational radiation reaction force experienced by the inner object. The mean motion resonance breaks down roughly at the point that gravitational radiation reaction exceeds the magnitude of the disk force.
Bottom left panel: the evolution of semi-major axes with respect to time.
Bottom right panel: the evolution of the eccentricities with respect to time.
}
\label{fig:alpha1}
\end{figure*}

\begin{figure}
\includegraphics[width=8.5cm]{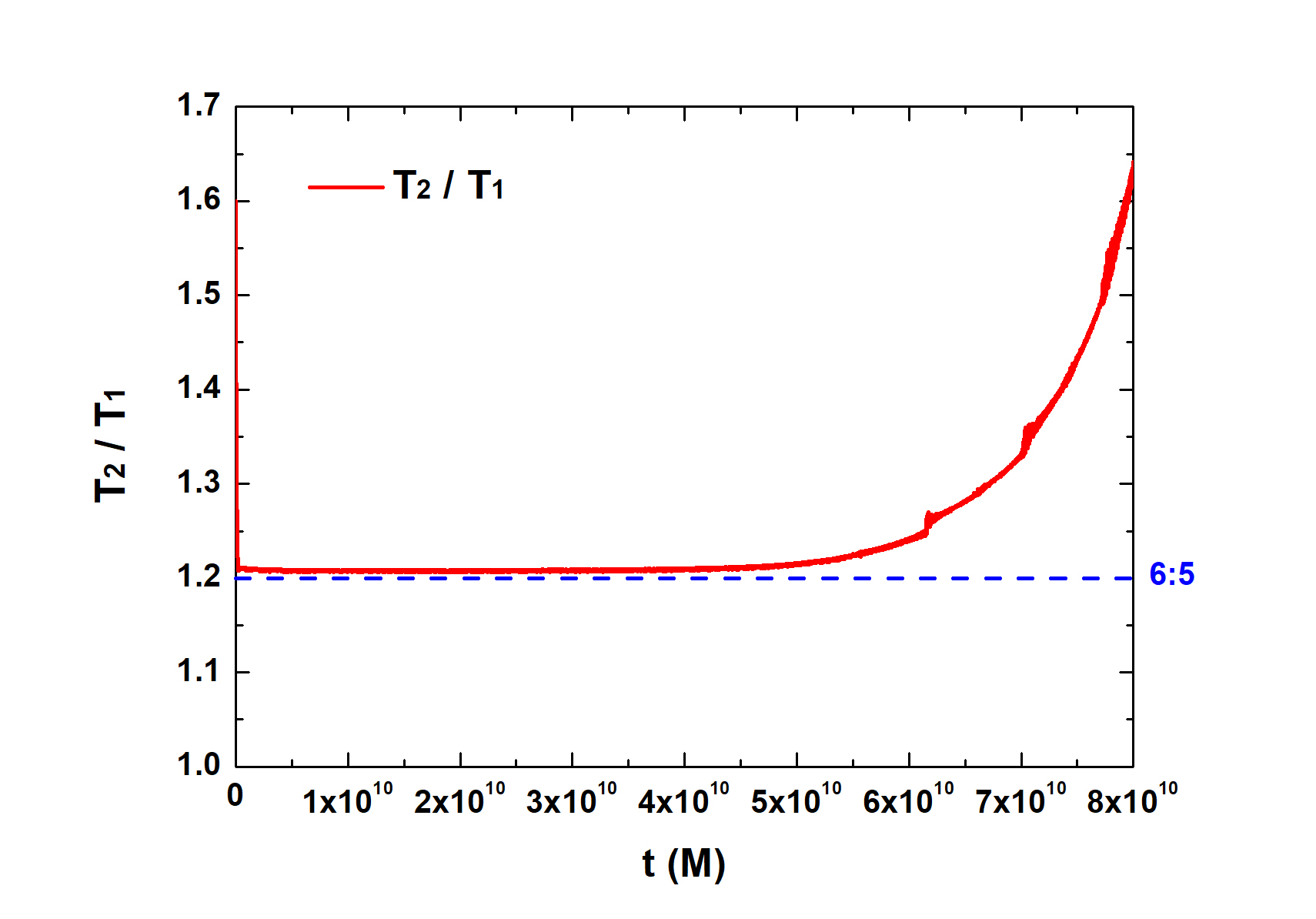}
\includegraphics[width=8.5cm]{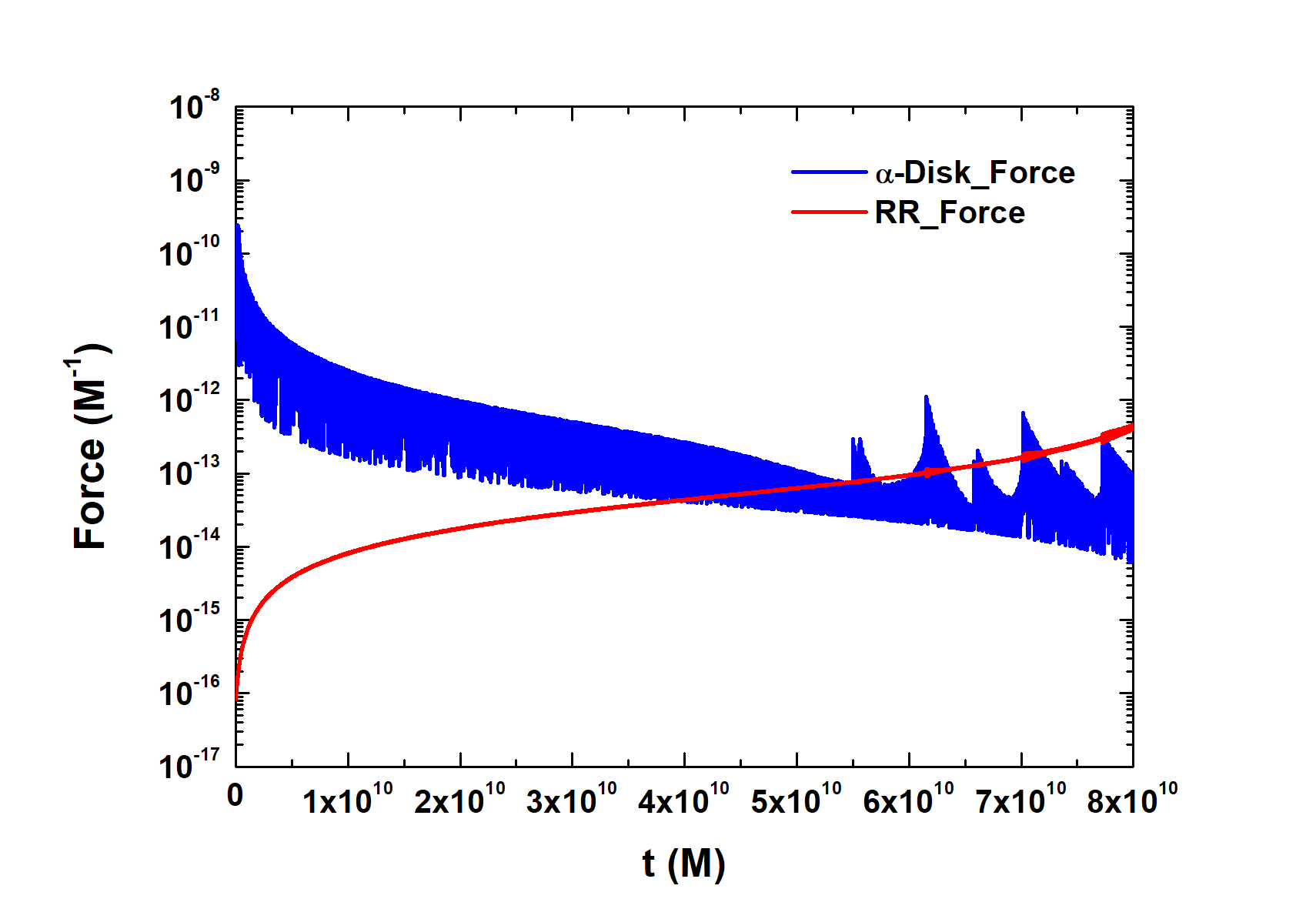}
\includegraphics[width=8.5cm]{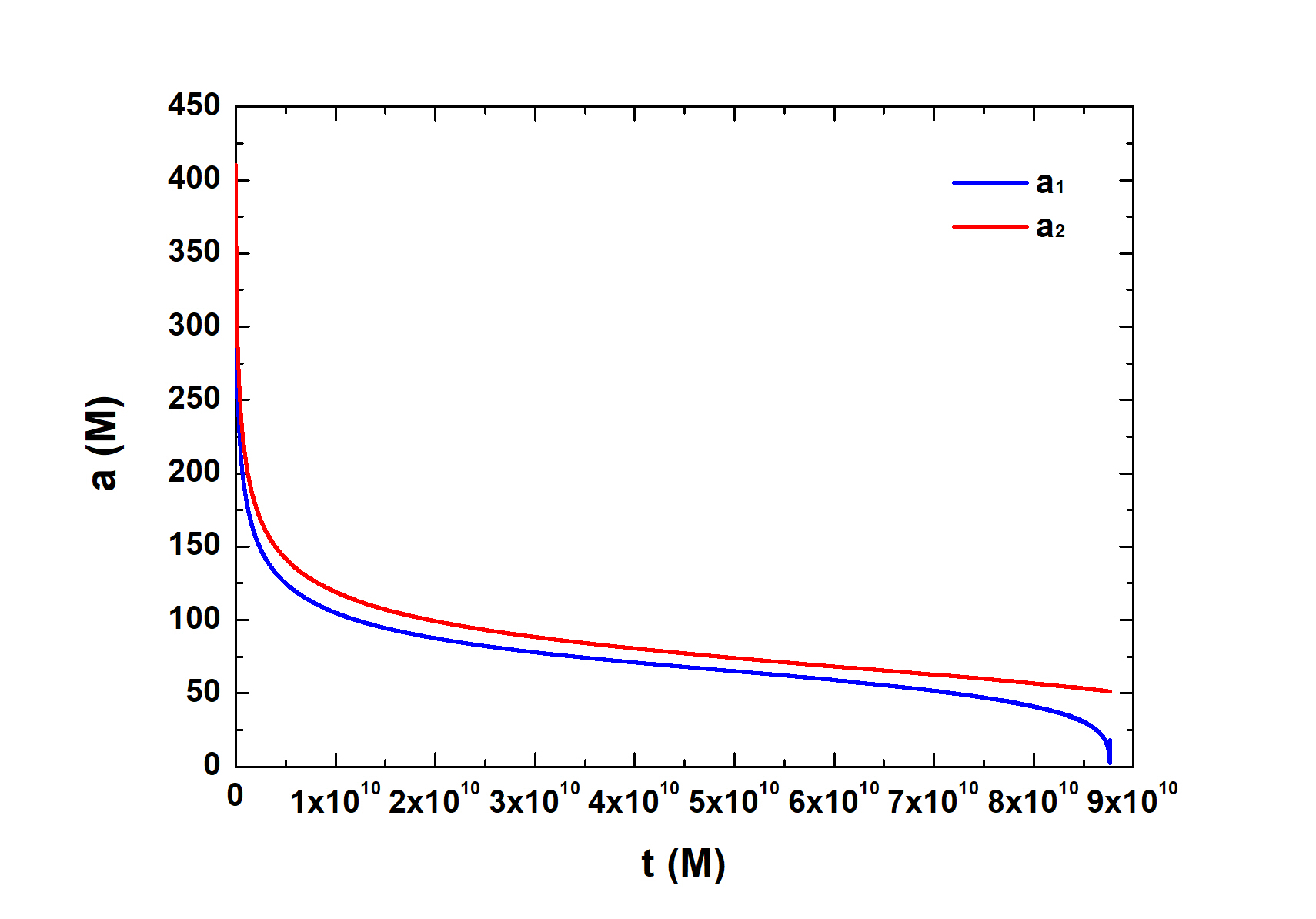}
\includegraphics[width=8.5cm]{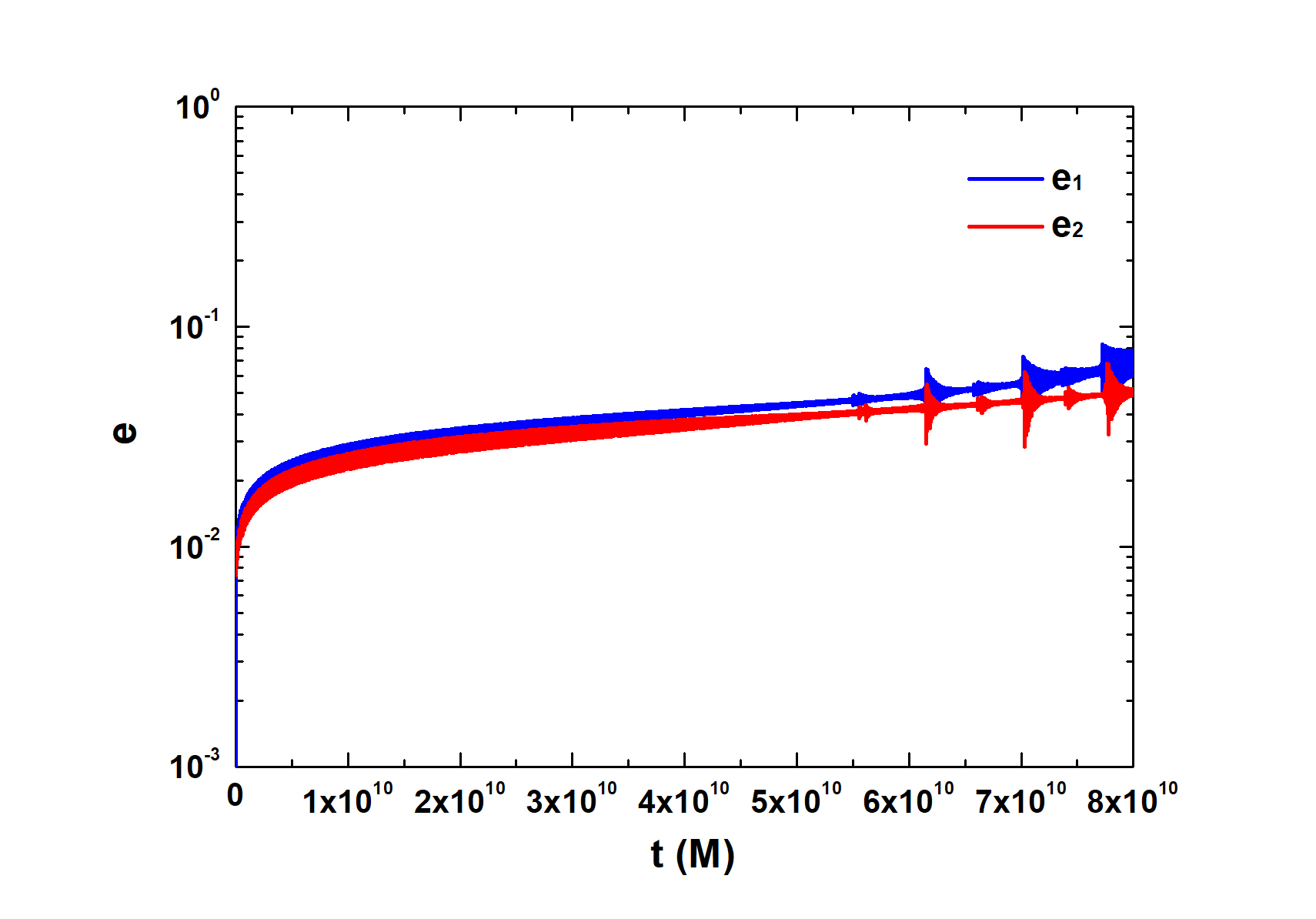}
\caption{Similar to Fig.~\ref{fig:alpha1}, except that the system starts with $a=300 M$ and  $\underline{a} =410.3742 M$ and later on is captured into $6:5$ resonance. It misses the $3:4$ resonance because at that point the migration rate is still too fast \cite{friedland2001migration}.}
\label{fig:alpha2}
\end{figure}

\begin{figure}
\includegraphics[width=8.5cm]{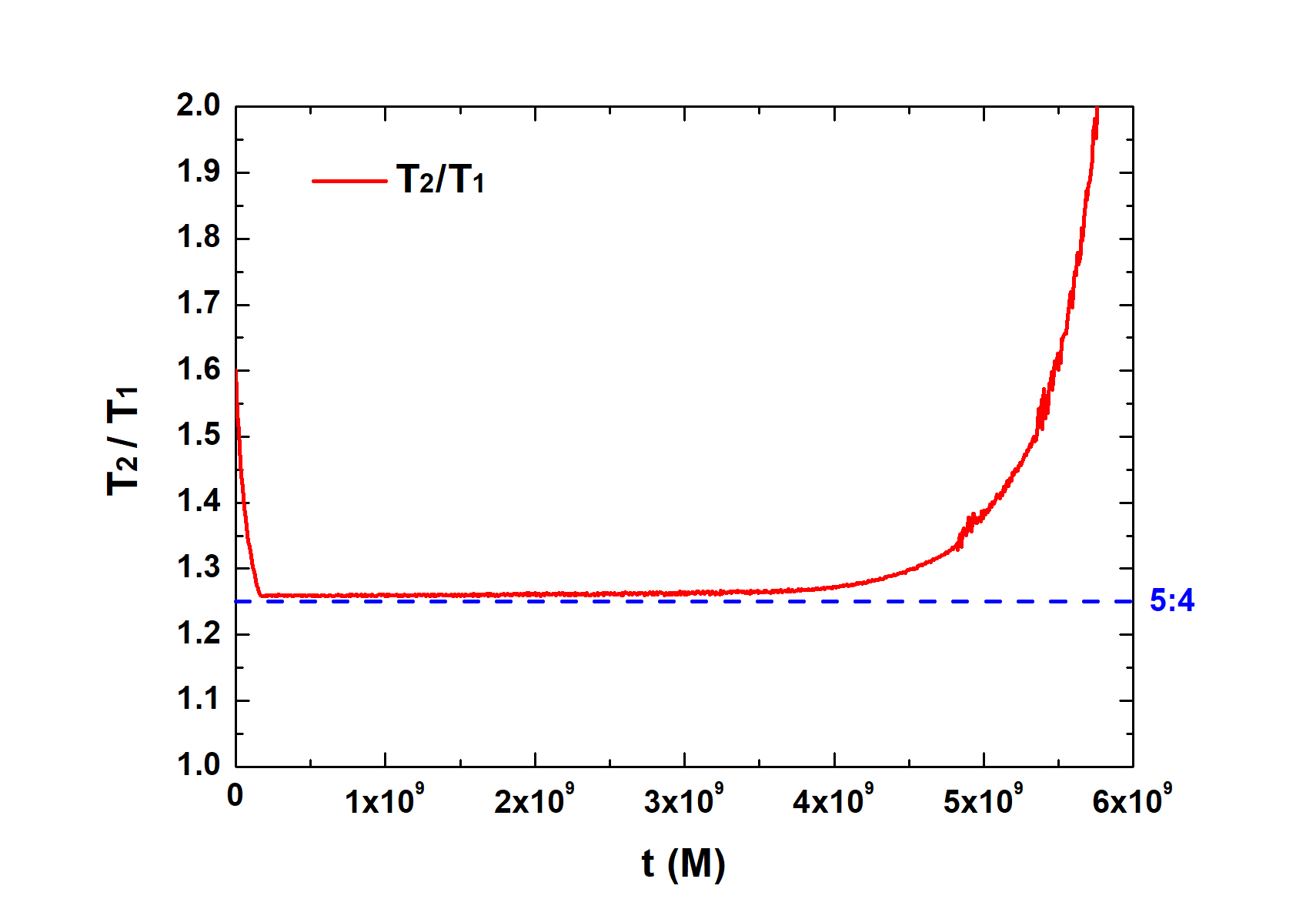}
\includegraphics[width=8.5cm]{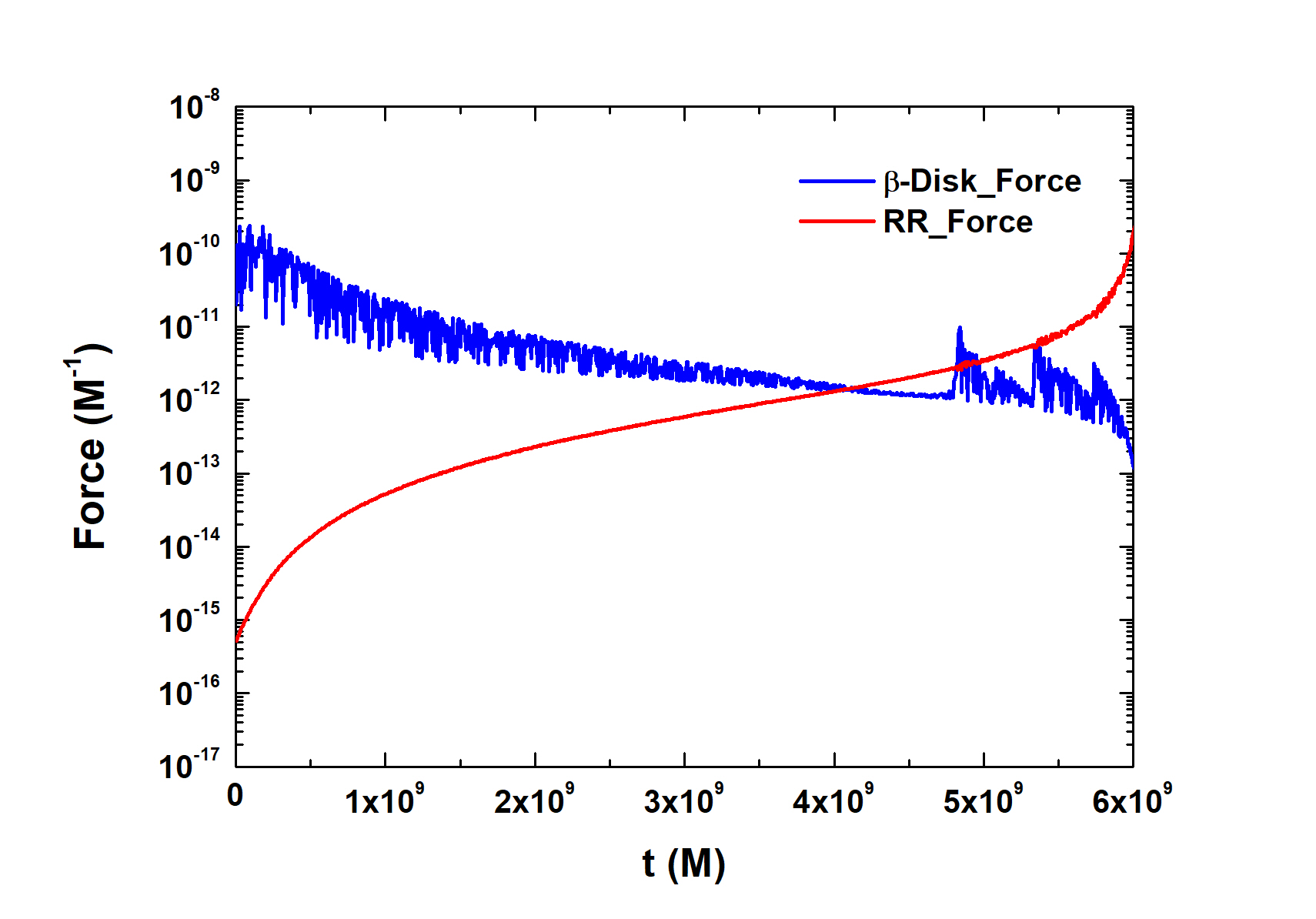}
\includegraphics[width=8.5cm]{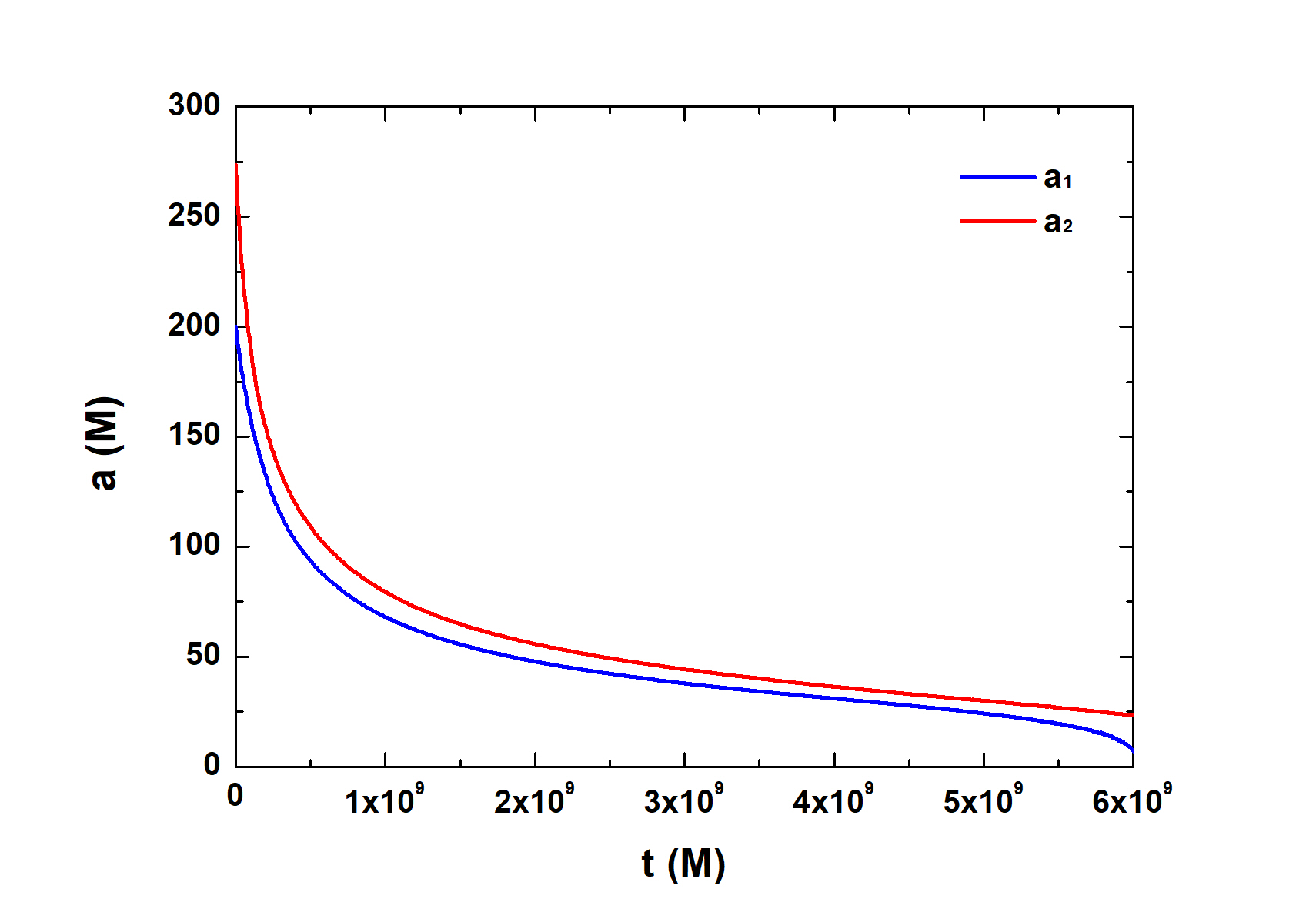}
\includegraphics[width=8.5cm]{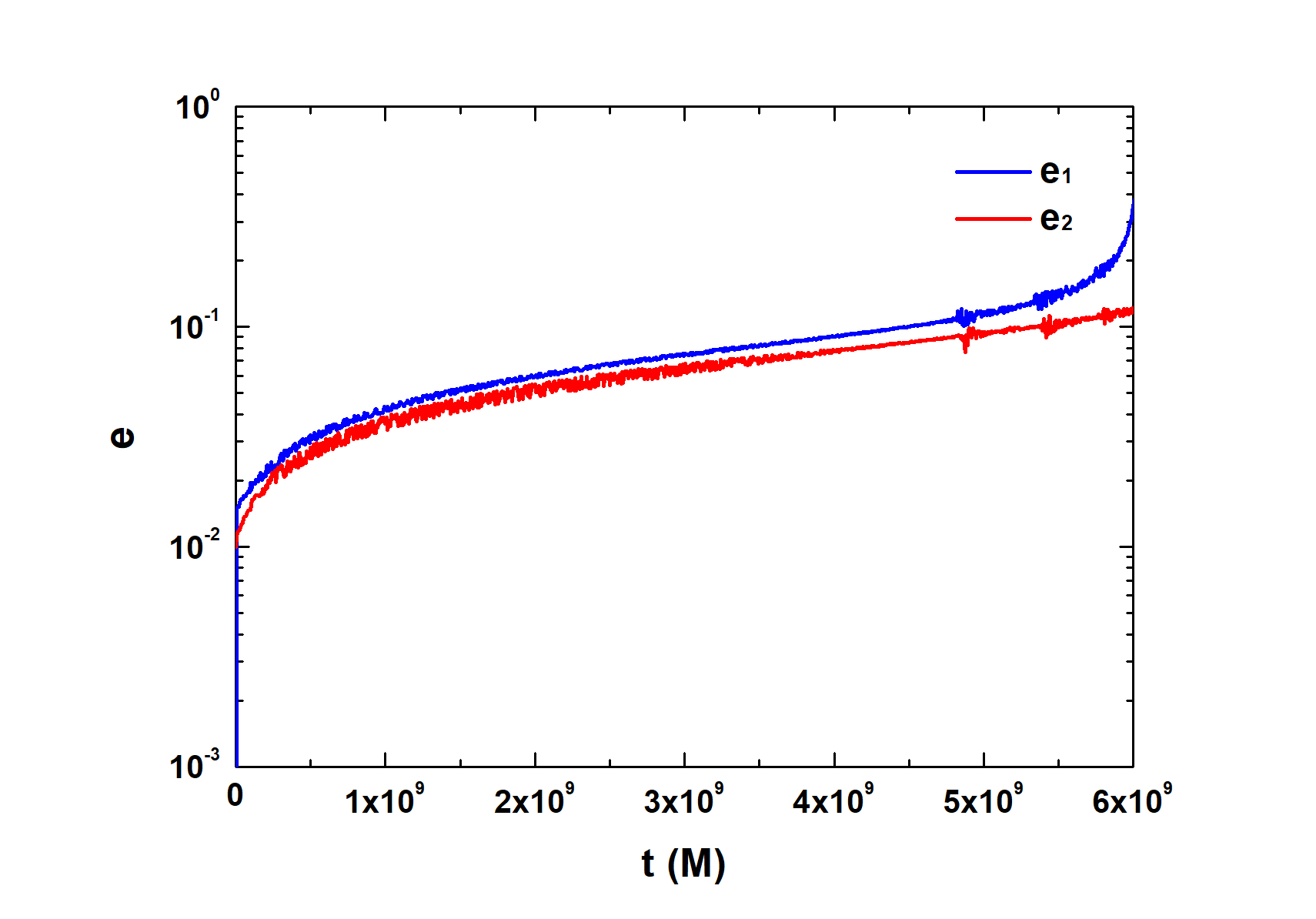}
\caption{Similar to Fig.~\ref{fig:alpha1}, except that the disk profile is a  $\beta$-disk and later on the system gets captured into $5:4$ resonance. It misses the $2:3$ and $3:4$ resonances because at those points the migration rate is still too fast \cite{friedland2001migration}.}
\label{fig:beta1}
\end{figure}

\begin{figure*}
\subfloat{
\includegraphics[width=8.5cm]{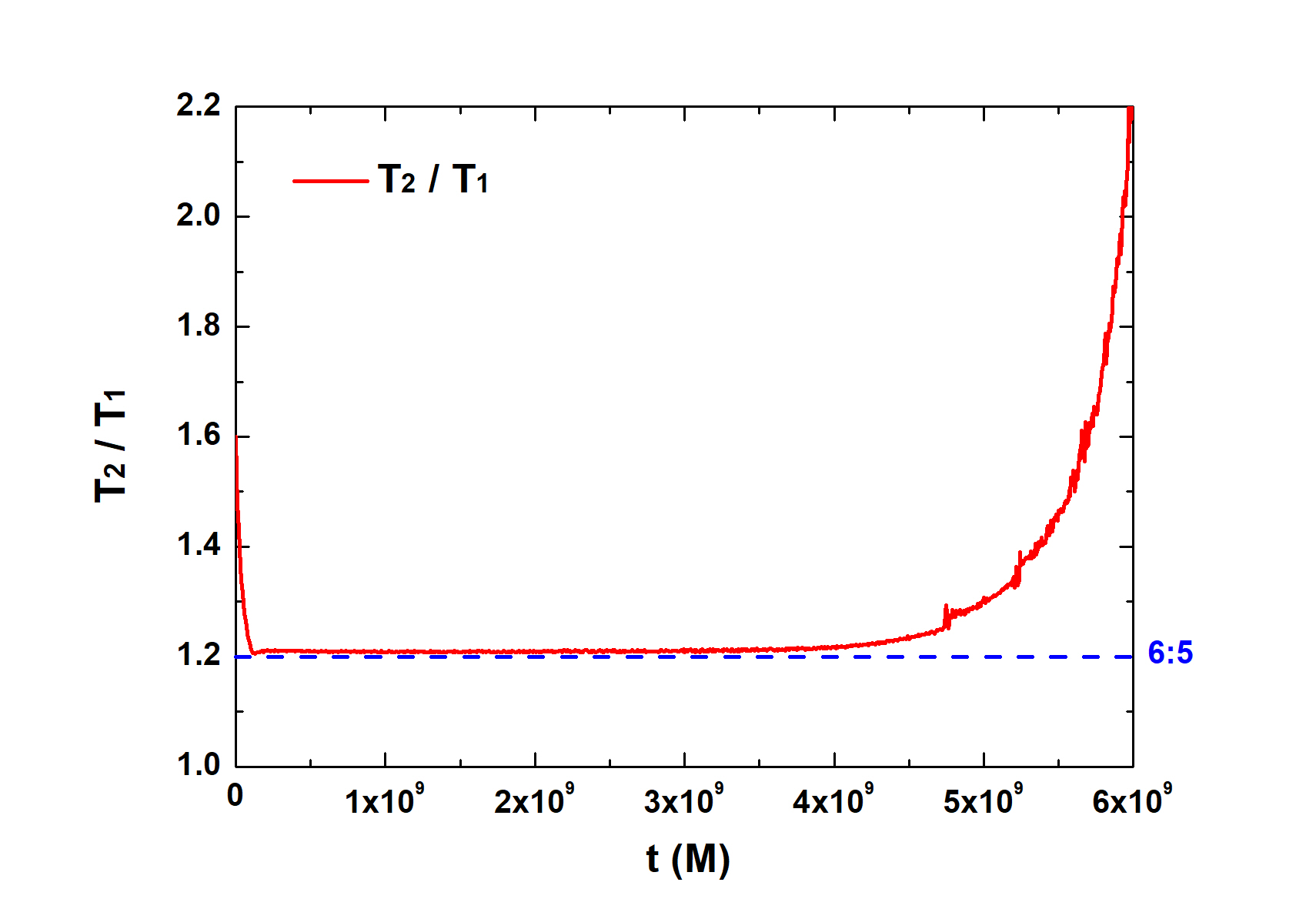}
}
\subfloat{
\includegraphics[width=8.5cm]{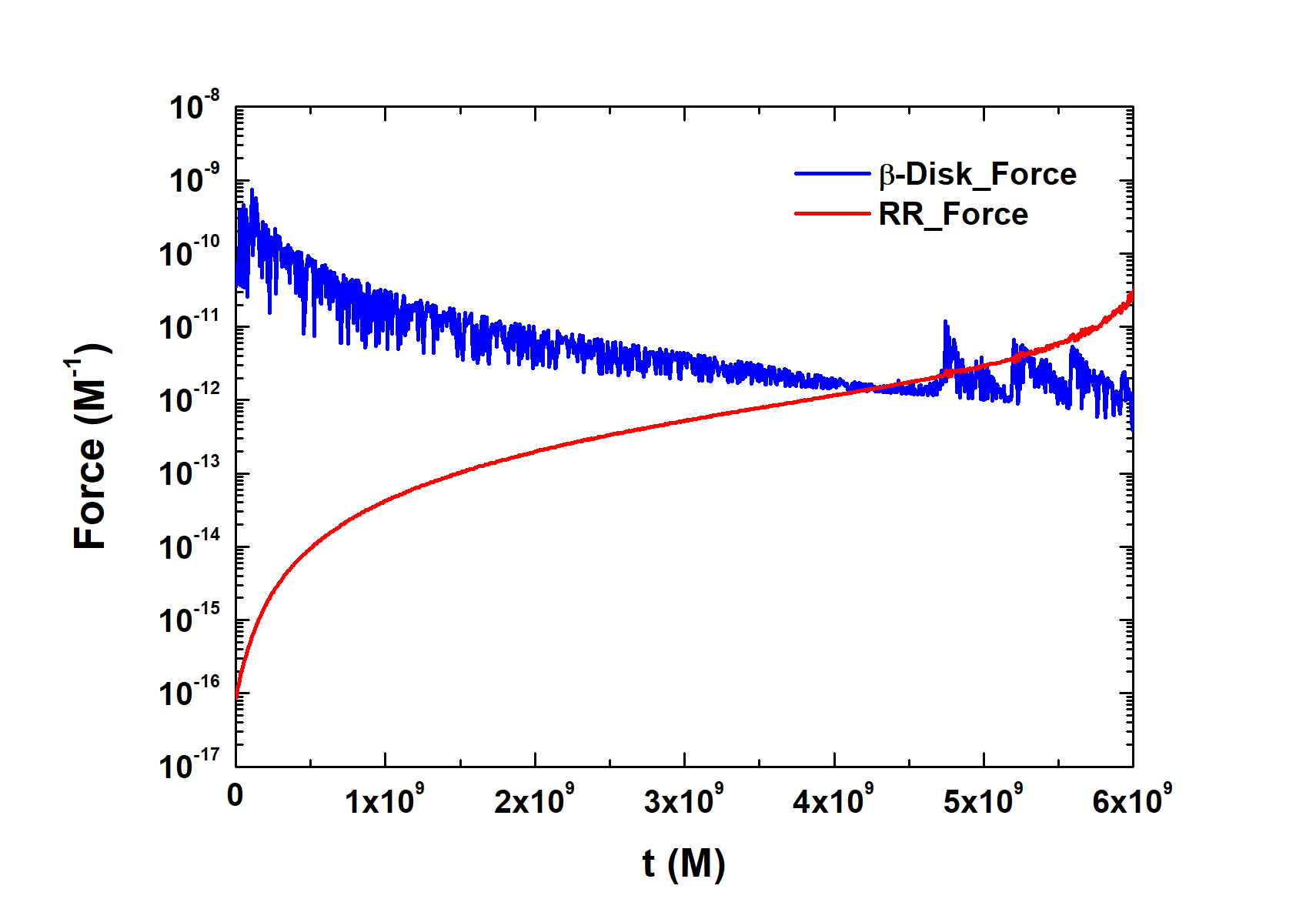}
}
\\
\subfloat{
\includegraphics[width=8.5cm]{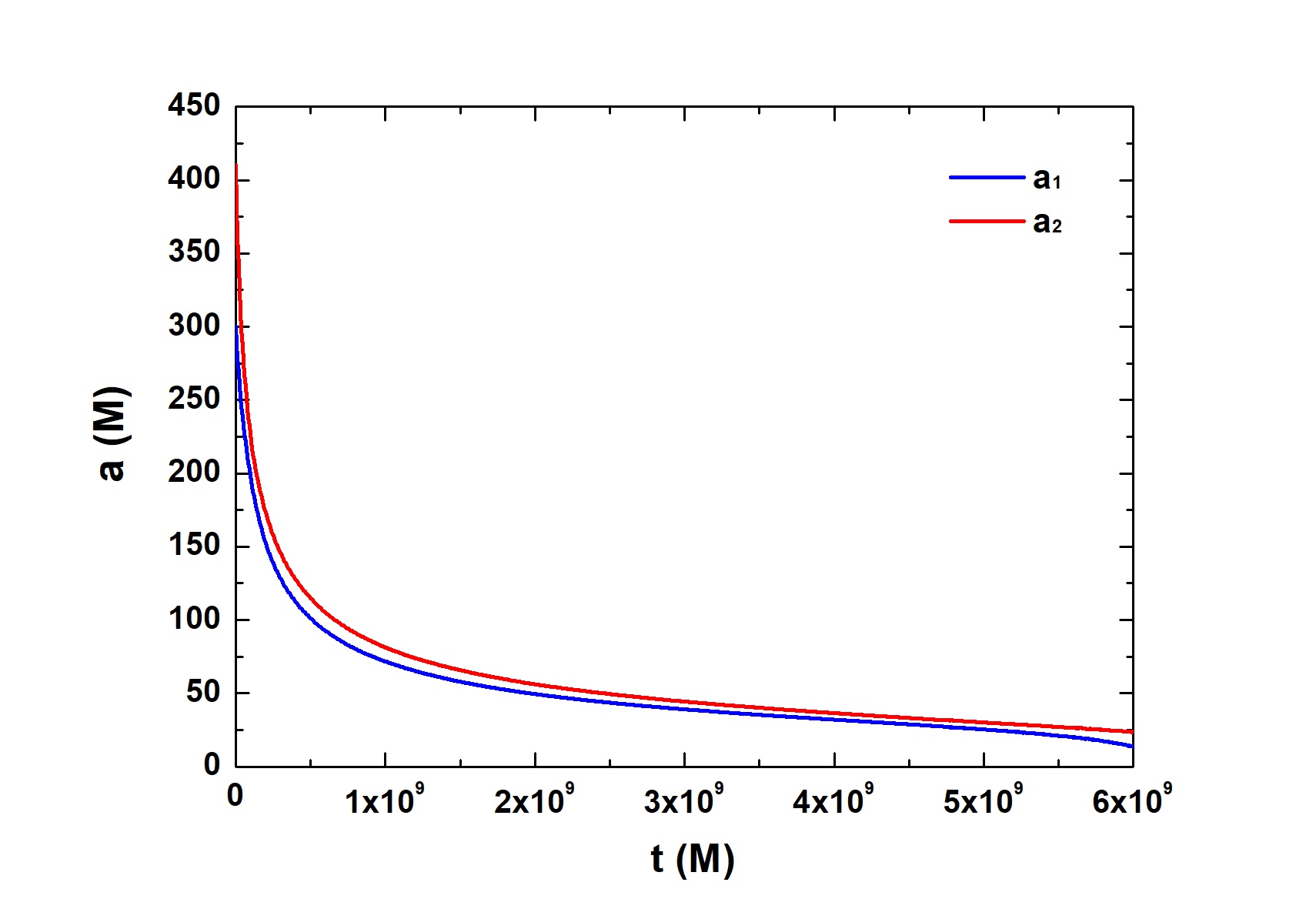}
}
\subfloat{
\includegraphics[width=8.5cm]{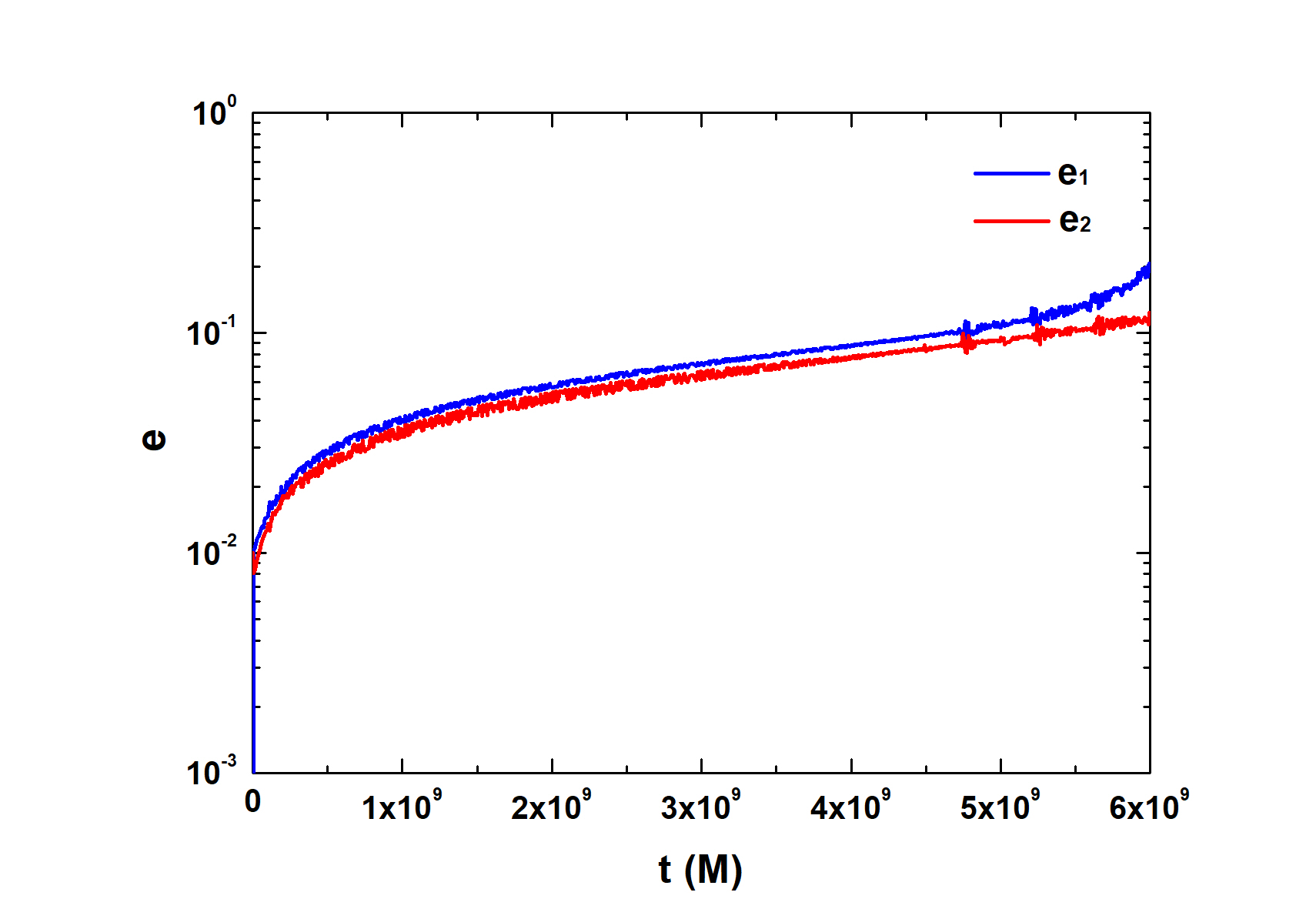}
}
\caption{Similar to Fig.~\ref{fig:alpha2}, except that the disk profile is a  $\beta$-disk and later on the system gets captured into $5:6$ resonance.It misses the $2:3$, $3:4$ and $4:5$ resonances because at those points the migration rate is still too fast \cite{friedland2001migration}.}
\label{fig:beta2}
\end{figure*}

In Fig.~\ref{fig:alpha1}-\ref{fig:beta2} we present the numerical evolution of two SBHs around a supermassive BH, with different initial separation and different disk models.
The exact form of the disk force is adapted from Section VII.B of \cite{kocsis2011observable} for the migration force and Section V. A of \cite{kocsis2011observable} for the accretion force.
The numerical evolution employs the N-body code REBOUND developed in \cite{rein2012rebound,rein2014ias15}, where we have added the leading order post-Newtonian corrections to the conservative and dissipative part of equation of motions and the disk force%
\footnote{There is an important caveat associated with the treatment of $\beta$-disks. In principle for the parameters assumed here, the SBHs may open gaps in the disk for $a \ge 100M$. The disk-SBH interaction will be stronger in the presence of a disk cavity. However, it is not clear how to obtain $\tau_e$ for an eccentric orbit with a disk cavity. We therefore still use the Type I disk force described in \cite{kocsis2011observable} for the numerical evolution.}.
In all these cases, the system is captured into a $j-1:j$ resonance until the point that the gravitational radiation reaction is greater than the disk force. During the mean motion resonance, the two SBHs migrate together towards the supermassive BH, while keeping the ratio of their periods $(j-1)/j$ roughly constant. 

In fact, this system is captured into an inner and outer resonance \textit{simultaneously}. The occurrence of pairs of resonances is not new and has been observed in other scenarios as well \cite{matsumoto2012orbital}.
Plots of the resonance angles demonstrating explicitly that the system is indeed captured into both an inner and outer resonance are shown in Appendix~\ref{sec:pair}, where we also included a short discussion on a subtle issue regarding the numerical extraction of these resonant angles when the eccentricities are small and post-Newtonian effects are important.

At the point where the mean motion resonance breaks, the outer object has already been brought to a rather close distance from the supermassive BH. While the inner SBH spirals into the supermassive BH and enters the LISA band, its motion will be affected by the gravitational field of the outer object. In \cite{yang2017general,bonga2019tidal} it has been shown that the main effect of the external perturber is to modify the angular momentum of the inner inspiraling binary, through an effect referred to as tidal resonance. This effect will be encoded into the gravitational radiation from the inner binary, which may be detected by LISA. It is also important to note that in general the disk rotation does not necessarily align with the spin of the supermassive black hole. In those cases we generally expect inclined extreme mass-ratio inspirals.

The period ratios shown in Fig.~\ref{fig:alpha1}-\ref{fig:beta2} are a few percent off the exact value $j:j-1$. Similar phenomena has been observed by the {\it Kepler} spacecraft \cite{fabrycky2014architecture} with many associated discussions in \cite{lithwick2012resonant,batygin2012dissipative,petrovich2013planets,baruteau2013disk}, although most of the Kepler systems are outside of the mean motion resonances.

\section{Conclusion}
\label{sec:conclusion}

In this work we consider the relativistic generalization of the mean motion resonance widely studied in planetary systems. The primary system of interest is a supermassive black hole with several stellar-mass black holes (SBHs) orbiting in its vicinity. Depending on the distance between the SBHs and the supermassive black hole, relativistic corrections may become important.

We have presented two separate analysis for this multi-body system, depending on the importance of relativistic correction. If the SBHs move within the strong gravity regime of the supermassive BH, the only reliable approach to describe their motion is black hole perturbation theory. In Sec.~\ref{sec:GR} we develop a Hamiltonian formalism using black hole perturbation theory, and find much richer structure for mean motion resonance in this fully relativistic regime. In fact, in general each SBH has three orbital frequencies for a Kerr geodesic motion, and the relativistic mean motion resonance could happen if the combination of six orbital frequencies of the two SBHs is zero.
Despite the theoretical interest in such resonance structure, it remains an open question which of these possible mean motion resonances is astrophysically relevant.

In the second approach, as discussed in Sec.~\ref{sec:post-newtonian}, we include post-Newtonian corrections to the equation of motion of the multi-body system.
This approach is physically more transparent, as the post-Newtonian Hamiltonian recovers the known Newtonian limit if we take $1/c^2 \rightarrow 0$.
It only applies for cases in which the post-Newtonian expansion is valid so that it is accurate to truncate the series after the first-order post-Newtonian terms, i.e. away from the strong-gravity regime. We find that the post-Newtonian correction does introduce the precession of the pericenters of the SBHs, but the qualitative structure of the resonances remains unchanged.
It is also worth to note that post-Newtonian effects in Kozai-Lidov mechanism have been studied in the past \cite{miller2002four,will2017orbital,naoz2013resonant}.

To illustrate possible way(s) to form resonant pairs of SBHs,  in Sec.\ref{sec:numerics} we have presented a few numerical examples for multi-SBHs moving within a thin accretion disk around a supermassive black hole.
In the regime that the disk force dominates over the gravitational radiation reaction, we observe sustained locking of the mean motion resonance, so that both SBHs migrate to close distances from the supermassive black hole until the gravitational radiation starts to dominate and the resonance breaks down. While this is a viable physical scenario, it remains an open question whether the conditions for this to occur are realized in astrophysical systems. Additionally, there may exist other astrophysical scenarios -- not explored here -- in which SBH pairs are locked into mean motion resonance within the gravitational influence sphere of a supermassive black hole.

If the resonance breaks down before the inner SBH enters the LISA band, the outer SBH will act as a gravitational perturber to the inner extreme mass-ratio inspiral within the LISA band. Such scenario has been discussed in \cite{bonga2019tidal}, where the main contribution from the outer SBH is through resonant kicks during tidal resonances \cite{yang2017general,bonga2019tidal}.  On the other hand, if the pair of SBHs is still locked into resonance once the inner SBH enters the LISA band, they must coherently spiral into the supermassive black hole, with the gravitational waveform vastly different from an ordinary extreme mass-ratio waveform.
According to the numerical examples studied in Sec.\ref{sec:numerics}, such a scenario probably only happens for $\beta$-disk models as these models have a much smaller resonance-breaking radius, or for possible SBHs surrounding intermediate mass black holes in dwarf galaxies \cite[e.g.,][]{Woo+2019}.

%{\bf comment on sustained transient resonance.}

\acknowledgments %
H. Y. and B.B. thank Eric Poisson and Cole Miller for interesting discussions.
This research was supported by NSERC and in part by the Perimeter
Institute for Theoretical Physics. Research at Perimeter Institute is
supported by the Government of Canada through the Department of
Innovation, Science and Economic Development Canada and by the
Province of Ontario through the Ministry of Research, Innovation and
Science.

\appendix

\section{Tensor harmonics}\label{sec:tensor}

We slightly modify the normalization of the original convention of Zerilli \cite{PhysRevD.2.2141} and notation in \cite{sago2003gauge} to evaluate metric perturbations in Sec.~\ref{sec:resonance-example}. The relevant tensor components are
\begin{equation}
{\bf a}^{(0)}_{\ell m} =
\left(\begin{array}{cccc} Y_{\ell m} & 0 & 0 & 0 \\
0 & 0 & 0 & 0 \\
0 & 0 & 0 & 0 \\
0 & 0 & 0 & 0
 \end{array}\right)\,,
 {\bf a}^{(1)}_{\ell m} =
\left(\begin{array}{cccc} 0 & Y_{\ell m} & 0 & 0 \\
Y_{\ell m} & 0 & 0 & 0 \\
0 & 0 & 0 & 0 \\
0 & 0 & 0 & 0
 \end{array}\right)\,,
\end{equation}

\begin{equation}
{\bf a}_{\ell m} =
\left(\begin{array}{cccc} 0 & 0 & 0 & 0 \\
0 & Y_{\ell m} & 0 & 0 \\
0 & 0 & 0 & 0 \\
0 & 0 & 0 & 0
 \end{array}\right)\,,
 {\bf b}^{(0)}_{\ell m} =
\left(\begin{array}{cccc} 0 & 0 &\frac{ \partial Y_{\ell m} }{\partial \theta}& \frac{ \partial Y_{\ell m}}{\partial \phi} \\
0 & 0 & 0 & 0 \\
\frac{ \partial Y_{\ell m} }{\partial \theta}  & 0 & 0 & 0 \\
\frac{ \partial Y_{\ell m}}{\partial \phi} & 0 & 0 & 0
 \end{array}\right)\,,
\end{equation}

\begin{equation}
{\bf c}^{(0)}_{\ell m} =
\left(\begin{array}{cccc} 0 & 0 & \frac{1}{\sin \theta} \frac{\partial Y_{\ell m}}{\partial \phi}  & -\sin\theta \frac{\partial Y_{\ell m}}{\partial \theta} \\
0 & 0 & 0 & 0 \\
\frac{1}{\sin \theta} \frac{\partial Y_{\ell m}}{\partial \phi} & 0 & 0 & 0 \\
-\sin\theta \frac{\partial Y_{\ell m}}{\partial \theta}& 0 & 0 & 0
 \end{array}\right)\,,
 \end{equation}
 \begin{equation}
 {\bf c}_{\ell m} =
\left(\begin{array}{cccc} 0 & 0 &0& 0 \\
0 & 0 & \frac{1}{\sin \theta} \frac{\partial Y_{\ell m}}{\partial \phi}  & -\sin\theta \frac{\partial Y_{\ell m}}{\partial \theta}\\
0  &   \frac{1}{\sin \theta} \frac{\partial Y_{\ell m}}{\partial \phi}  & 0 & 0 \\
0 &  -\sin\theta \frac{\partial Y_{\ell m}}{\partial \theta} & 0 & 0
 \end{array}\right)\,,
\end{equation}

\begin{equation}
{\bf d}_{\ell m} =
\left(\begin{array}{cccc} 0 & 0 & 0 & 0 \\
0 & 0 & 0 & 0 \\
0 & 0 & -\frac{X_{\ell m}}{\sin \theta} &  \sin \theta W_{\ell m} \\
0 & 0 & \sin \theta W_{\ell m} & \sin \theta X_{\ell m}
 \end{array}\right)\,,
 \end{equation}
 \begin{equation}
 {\bf g}_{\ell m} =
\left(\begin{array}{cccc} 0 & 0 & 0 & 0 \\
0 & 0 & 0 & 0 \\
0 & 0 & Y_{\ell m} & 0 \\
0 & 0 & 0 & \sin^2\theta Y_{\ell m}
 \end{array}\right)\,,
\end{equation}
\begin{equation}
 {\bf f}_{\ell m} =
\left(\begin{array}{cccc} 0 & 0 & 0 & 0 \\
0 & 0 & 0 & 0 \\
0 & 0 & W_{\ell m} & X_{\ell m} \\
0 & 0 & X_{\ell m} & -\sin^2\theta W_{\ell m}
 \end{array}\right)\,.
\end{equation}
The tensor harmonic functions are
\begin{align}
& X_{\ell m} = 2 \frac{\partial }{ \partial \phi} \left ( \frac{\partial}{\partial \theta} -\cot \theta \right ) Y_{\ell m}\,, \nonumber \\
& W_{\ell m} = \left ( \frac{\partial^2}{\partial \theta^2} -\cot\theta \frac{\partial}{\partial \theta} -\frac{1}{\sin^2\theta} \frac{\partial^2}{\partial \phi^2}\right ) Y_{\ell m} \,.
\end{align}

\section{Resonant angles} \label{sec:pair}
\begin{figure*}
\subfloat{
\includegraphics[width=8.5cm]{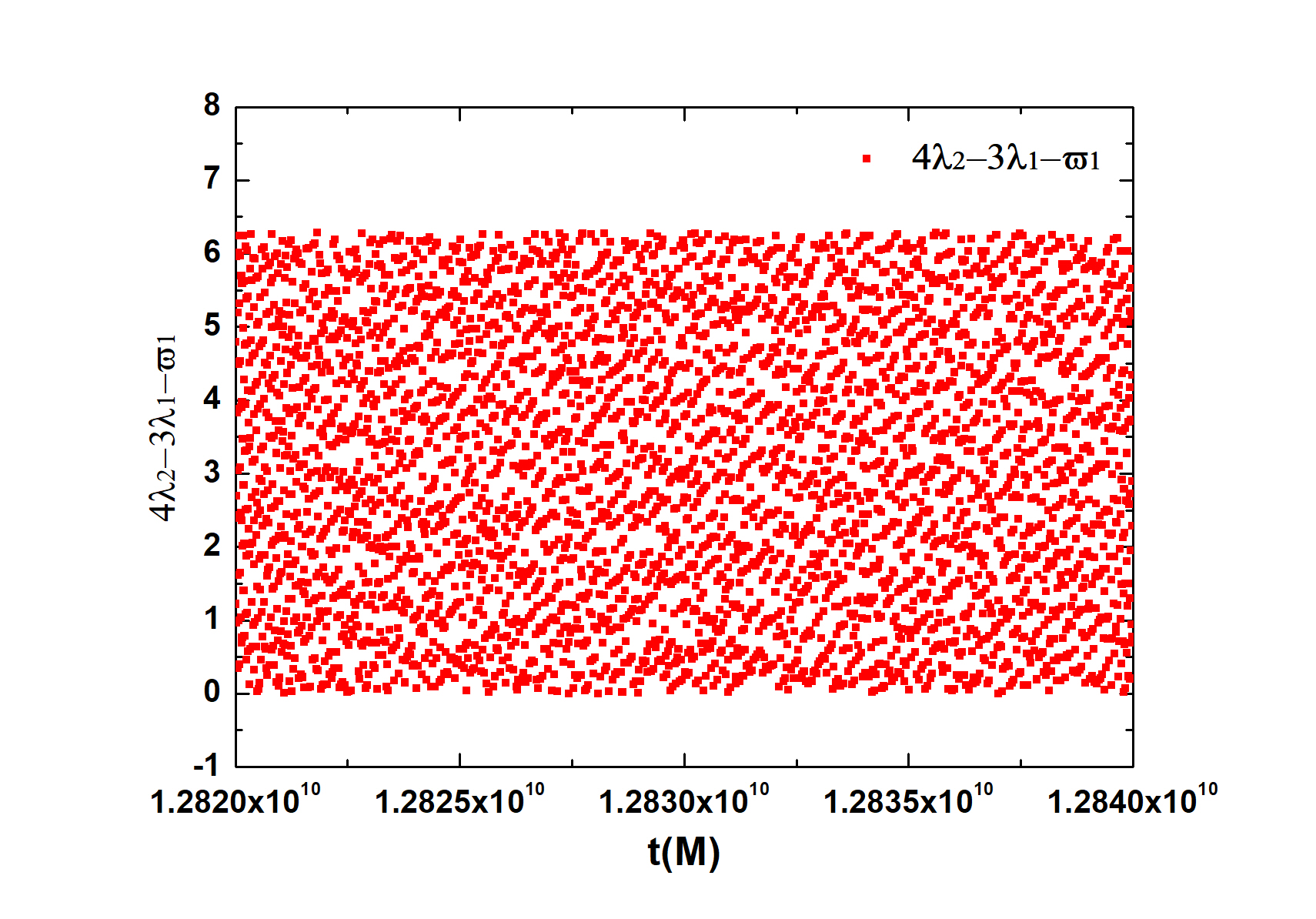}
}
\subfloat{
\includegraphics[width=8.5cm]{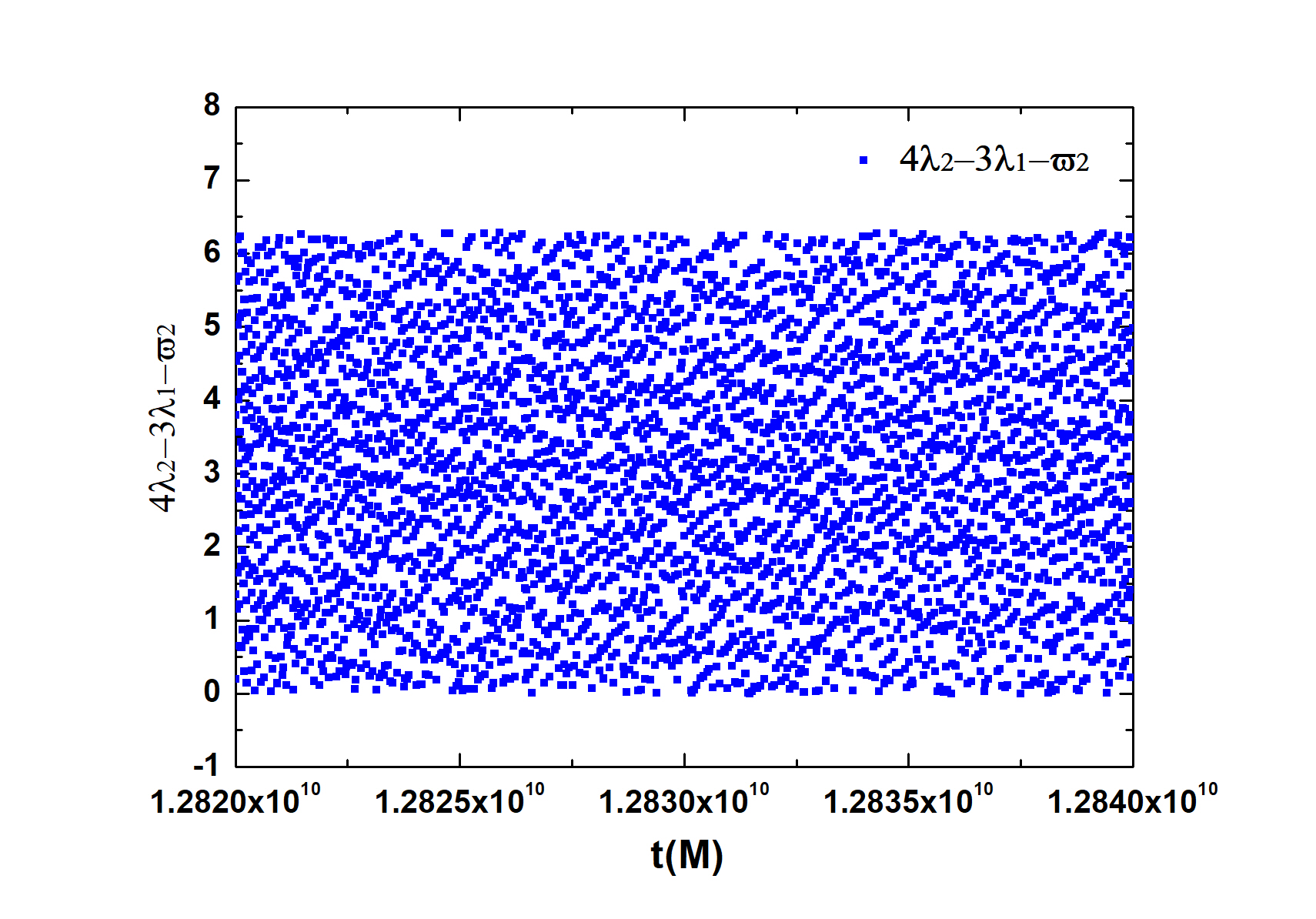}
}
\\
\subfloat{
\includegraphics[width=8.5cm]{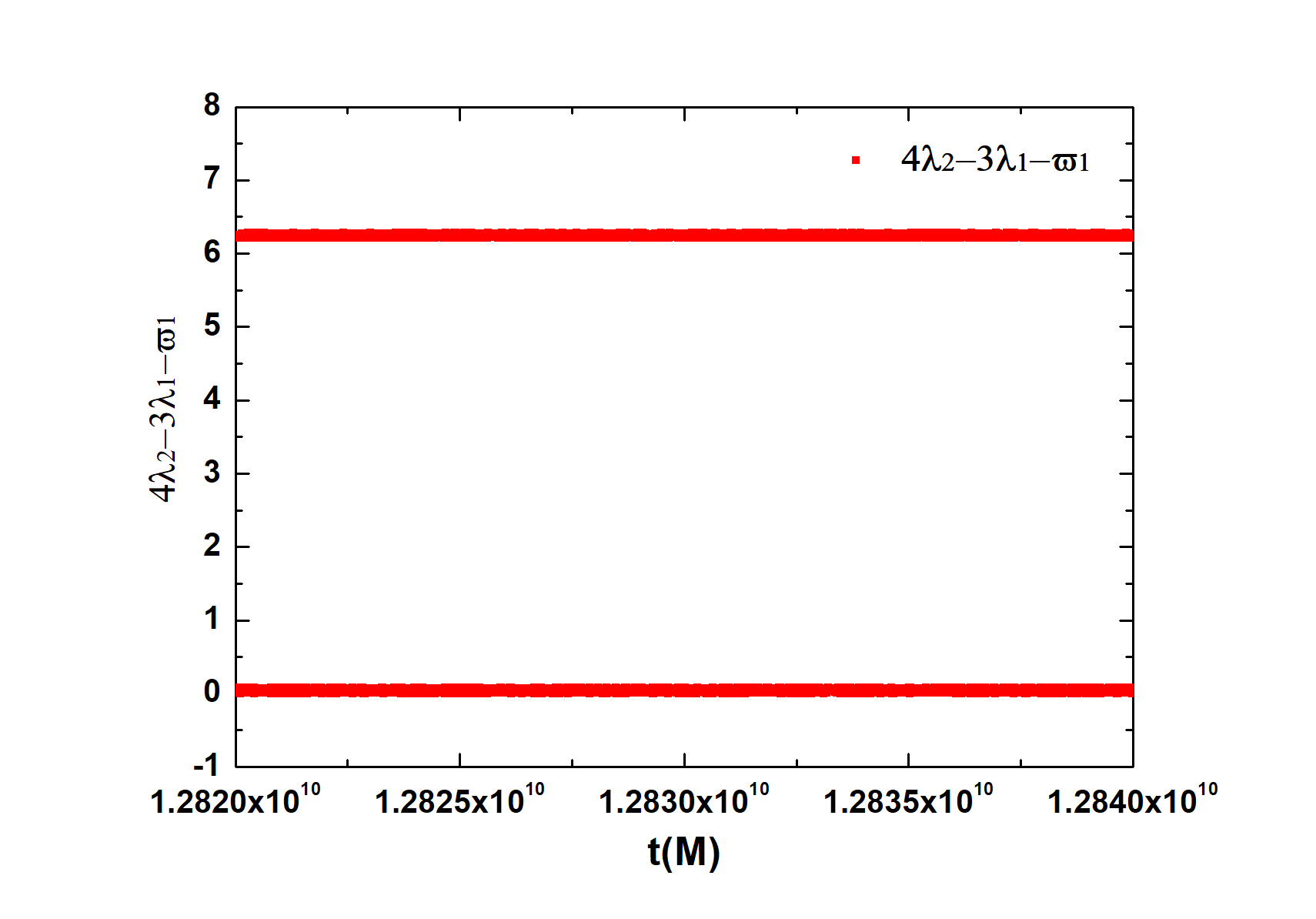}
}
\subfloat{
\includegraphics[width=8.5cm]{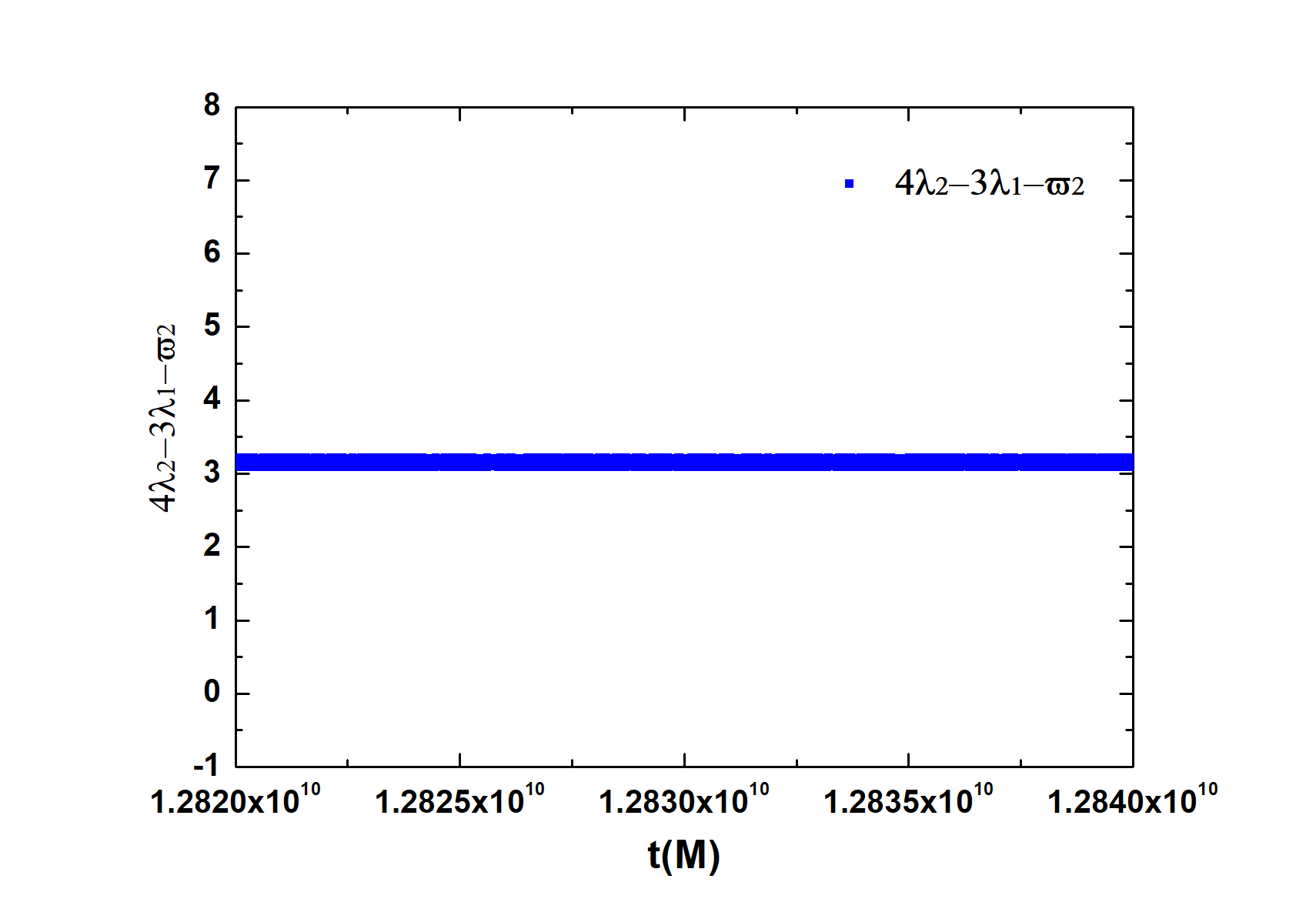}
}
\caption{Top left/right panel: Resonant angles $\theta_{1}$/$\theta_2$ of the osculating orbit corresponding to the simulation shown in Fig.~\ref{fig:alpha1} during a representative time frame after the ratio of the period is roughly $4:3$. We do not observe locking of the resonant angles. 
Bottom left/right panel: Removing the 1PN term that generates the general relativistic precession, resonant angle locking is restored. 
(In order to be consistent to the literature in planetary science \cite{Murray-Dermott}, we denote the argument of pericentre as $\varpi$, which is just $-\gamma$ in Sec.\ref{sec:post-newtonian}.)}
\label{fig:alpha2angle}
\end{figure*}

In order to verify that a given system indeed resides in one (or more) mean motion resonances, it is important to check whether the corresponding resonant angles $\theta_{1,2}$ librate around a constant value. 
Surprisingly, even when the period ratios are clearly locked near $j-1 :j$ in Fig.~\ref{fig:alpha1}-\ref{fig:beta2}, there is no sign of resonant angle locking given an instantaneous extraction of the resonant angles. Interestingly, if we remove the first post-Newtonian Hamiltonian in the equation of motion and perform the simulation again, the resonant angle locking is clearly restored, as shown in Fig.~\ref{fig:alpha2angle}. Naively, this seems to suggest that post-Newtonian corrections prohibit the system from entering into mean motion resonance. This is not correct: the system does experience mean motion resonance. The resolution is provided by how the resonant angles are extracted from the data.

It turns out that it has been long known that post-Newtonian corrections to the equations of motion may give rise to ``perpetual precession'' when the eccentricity is small \cite{lincoln1990coalescing,will2019compact}%
\footnote{This effect is not unique to post-Newtonian theory and other ``strange'' behaviors when eccentricities are small have also been studied in the planetary science community, see \cite{will2019compact} for references.}.
In other words, for small eccentricities the precession rate induced by the post-Newtonian corrections becomes the same as the orbital frequency, and as a result the true anomaly stays roughly constant in the post-Newtonian osculating description. In such cases, the physical orbits can be circular even if the osculating orbit is eccentric. This is exactly what is happening here. 
From the data we extract from REBOUND where the Poincar\'e variables are obtained by fitting instantaneous motion by elliptical orbits (the ``osculating orbit'' approximation), we do observe that the true anomaly stays approximately constant ($\sim \pi$) and the precession rate is the same as the orbital frequency. 
On the other hand, the physically eccentricity --- measured by comparing the maximum and minimum distance from the supermassive black hole on orbital timescales --- is on the order of $10^{-3}$ in contrast to the osculating eccentricity that is on the order of $10^{-2}$. 

\begin{figure*}
\subfloat{
\includegraphics[width=8.5cm]{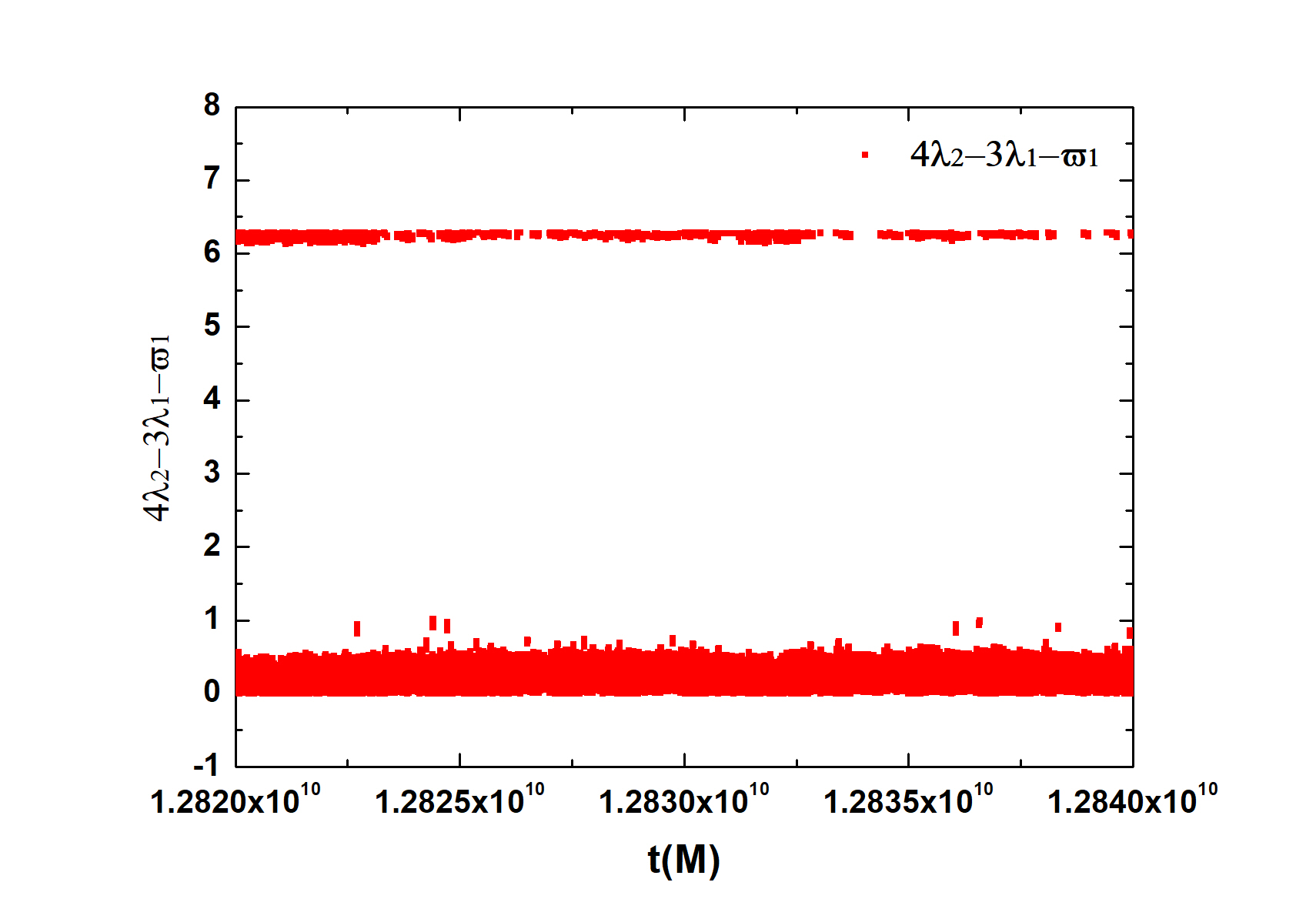}
}
\subfloat{
\includegraphics[width=8.5cm]{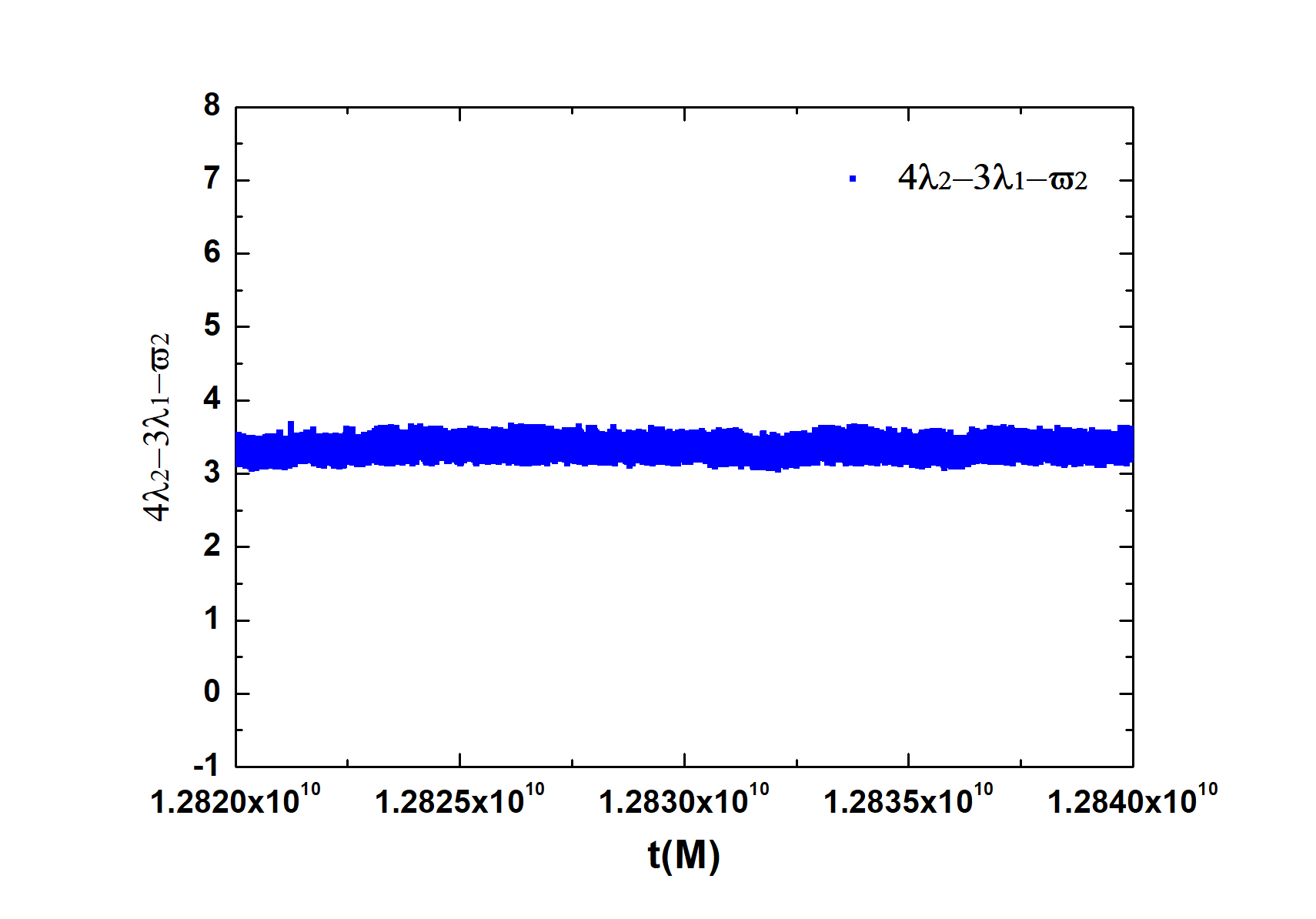}
}
\caption{Resonant angles $\theta_{1}$ and $\theta_2$ of the physical orbit corresponding to the simulation shown in Fig.~\ref{fig:alpha1}. The resonant angles are both locked in this case.}
\label{fig:phy}
\end{figure*}

In Fig.~\ref{fig:phy} we drop the osculating orbit assumption and extract the angles from the physical orbit. The corresponding resonant angles of the physical orbits are clearly locked, also when post-Newtonian corrections are included. This also explains why the period ratio stays constant during resonance.

%%%%%%%%%%%%%%%%%%%%%%%%%%%%%%%%%%%%%%%%%%%%
%%%%%%%%%%%%%%%%%%%%%%%%%%%%%%%%%%%%%%%%%%%%
\bibliography{References}
%\bibliography{master}
\end{document}